\pdfoutput=1
\newcommand*{\ATLASLATEXPATH}{latex/}
\documentclass[UKenglish,texlive=2016,cernpreprint,txfonts]{\ATLASLATEXPATH atlasdoc}
\usepackage[subcaption, biblatex=true]{\ATLASLATEXPATH atlaspackage}
\usepackage{\ATLASLATEXPATH atlasbiblatex}
\usepackage{\ATLASLATEXPATH atlascontribute}
\usepackage{\ATLASLATEXPATH atlasmisc}
\usepackage{\ATLASLATEXPATH atlasphysics}

\graphicspath{{logos/}{figures/}}

\usepackage{ss3l2017pub-defs}

\AtlasTitle{Search for supersymmetry in final states with two same-sign or three leptons and jets using 36 fb$^{-1}$ of 
$\sqrt{s}=13$~TeV $pp$ collision data with the ATLAS detector}

\author{The ATLAS Collaboration}

\AtlasRefCode{SUSY-2016-14}
\PreprintIdNumber{CERN-EP-2017-108}
\arXivId{1706.03731}
\HepDataRecord{77719}
\AtlasJournalRef{JHEP 09 (2017) 084}
\AtlasDOI{10.1007/JHEP09(2017)084}

\AtlasAbstract{%

A search for strongly produced supersymmetric particles using signatures involving multiple energetic jets and either two isolated 
same-sign leptons ($e$ or $\mu$), or at least three isolated leptons, is presented. 
The analysis relies on the identification of $b$-jets and high missing transverse momentum to achieve good sensitivity. 
A data sample of proton--proton collisions at $\sqrt{s}= 13$~TeV recorded with the ATLAS detector at the Large Hadron Collider in 
2015 and 2016, corresponding to a total integrated luminosity of 36.1 fb$^{-1}$, is used for the search. 
No significant excess over the Standard Model prediction is observed. The results are interpreted in several 
simplified supersymmetric models featuring $R$-parity conservation or $R$-parity violation, extending the exclusion 
limits from previous searches. In models considering gluino pair production, gluino masses are excluded up to 
1.87 TeV at 95\% confidence level. When bottom squarks are pair-produced and decay to a chargino and a top quark, models 
with bottom squark masses below 700 GeV and light neutralinos are excluded at 95\% confidence level. 
In addition, model-independent limits are set on a possible contribution of new phenomena to the signal region yields.
}

\hypersetup{pdftitle={ATLAS document},pdfauthor={The ATLAS Collaboration}}

\begin{document}

\maketitle


\section{Introduction}
\label{sec:intro}

Supersymmetry (SUSY)~\cite{Golfand:1971iw,Volkov:1973ix,Wess:1974tw,Wess:1974jb,Ferrara:1974pu,Salam:1974ig} is one of the best-motivated
extensions of the Standard Model (SM). A general review can be found in Ref.~\cite{Martin:1997ns}. In its minimal realization 
(the MSSM)~\cite{Fayet:1976et,Fayet:1977yc} it predicts a new bosonic (fermionic) partner for each fundamental SM fermion (boson), 
as well as an additional Higgs doublet. If $R$-parity~\cite{Farrar:1978xj} is conserved (RPC) the lightest supersymmetric particle (LSP) 
is stable and can be the lightest neutralino\footnote{The SUSY partners of the Higgs and electroweak gauge bosons, the
electroweakinos, mix to form
the mass eigenstates known as charginos ($\tilde{\chi}^{\pm}_{l}$, $l = 1, 2$ ordered by increasing mass) and neutralinos 
($\tilde{\chi}^{0}_{m}$, $m = 1, \ldots, 4$ ordered by increasing mass).} \ninoone. 
In many models, the LSP can be a dark-matter candidate~\cite{Goldberg:1983nd,Ellis:1983ew} and produce signatures with large missing 
transverse momentum. If instead $R$-parity is violated (RPV), the LSP decay can generate events with high jet and lepton multiplicity. 
Both RPC and RPV scenarios can produce the final-state signatures considered in this article.

In order to address the SM hierarchy problem with SUSY 
models~\cite{Sakai:1981gr,Dimopoulos:1981yj,Ibanez:1981yh,Dimopoulos:1981zb}, 
TeV-scale masses are required~\cite{Barbieri:1987fn,deCarlos:1993yy} for the partners of the gluons (gluinos~$\tilde{g}$) 
and of the top quarks (top squarks \stopL and \stopR), due to the large top Yukawa coupling.\footnote{The 
partners of the left-handed (right-handed) quarks are labelled $\tilde{q}_{\textrm{L(R)}}$. In the case where there is significant L/R 
mixing (as is the case for third-generation squarks) the mass eigenstates of these squarks are labelled $\tilde{q}_{1,2}$ ordered 
by increasing mass.} The latter also favours significant $\stopL$--$\stopR$ mixing, 
so that the mass eigenstate $\stopone$ is lighter than all the other squarks in many scenarios~\cite{Inoue:1982pi,Ellis:1983ed}. 
Bottom squarks ($\tilde{b}^{}_1$) may also be light, being bound to top squarks by $SU(2)_{\textrm{L}}$ invariance. 
This leads to potentially large production cross-sections 
for gluino pairs ($\gluino\gluino$), top--antitop squark pairs ($\stopone\stoponebar$) and bottom--antibottom squark pairs 
($\sbottomone\sbottomonebar$) at the Large Hadron Collider (LHC)~\cite{Borschensky:2014cia}. 
Production of isolated leptons may arise in the cascade decays of those superpartners to SM quarks and neutralinos~\ninoone, 
via intermediate neutralinos~$\tilde\chi^0_{2,3,4}$ or charginos~$\tilde\chi^\pm_{1,2}$ 
that in turn lead to $W$, $Z$ or Higgs bosons, or to lepton superpartners (sleptons, $\tilde{\ell}$). 
Light third-generation squarks would also enhance gluino decays to top or bottom quarks 
relative to the generic decays involving light-flavour squarks,
favouring the production of heavy-flavour quarks and, in the case of top quarks, additional isolated leptons. 

This article presents a search for SUSY in final states with two leptons (electrons or muons) of the same electric charge, 
referred to as same-sign (SS) leptons, or three leptons (3L), jets and in some cases also missing transverse momentum,
whose magnitude is referred to as \met. Only prompt decays of SUSY particles are considered.
It is an extension of an earlier search performed by the ATLAS experiment~\cite{PERF-2007-01} with $\sqrt s=13$ TeV data~\cite{paperSS3L}, 
and uses the data collected in proton--proton ($pp$) collisions during 2015 and 2016.  
Similar searches for SUSY in this topology were also performed by the CMS experiment at $\sqrt s=13$~TeV~\cite{Khachatryan:2016kod, 
Khachatryan:2017qgo,Sirunyan:2017uyt}. While the same-sign or three-lepton signatures are present in many scenarios of physics beyond the SM (BSM), 
SM processes leading to such final states have very small cross-sections. 
Compared to other BSM searches, analyses based on these signatures therefore allow 
the use of looser kinematic requirements (for example, on \met or on the momentum of jets and leptons), 
preserving sensitivity to scenarios with small mass differences between the produced gluinos/squarks and the LSP, or in which 
$R$-parity is not conserved. This sensitivity to a wide range of BSM physics processes is illustrated by the interpretation of the 
results in the context of twelve different SUSY simplified models~\cite{Alwall:2008ve,Alwall:2008ag,Alves:2011wf} 
that may lead to same-sign or three-lepton signatures.

\begin{figure}[t!]
\centering
\begin{tabular}{rrrr}
\begin{subfigure}[t]{0.24\textwidth}\includegraphics[width=\textwidth]{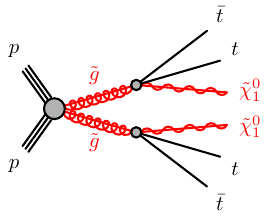}\caption{}\label{fig:feynman_gtt}\end{subfigure} &
\begin{subfigure}[t]{0.24\textwidth}\includegraphics[width=\textwidth]{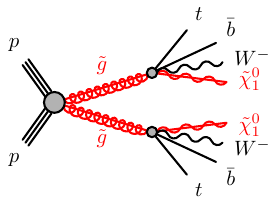}\caption{}\label{fig:feynman_gttOffshell}\end{subfigure} &
\begin{subfigure}[t]{0.24\textwidth}\includegraphics[width=\textwidth]{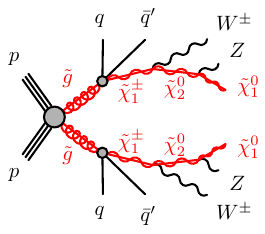}\caption{}\label{fig:feynman_gg2WZ}\end{subfigure} &
\begin{subfigure}[t]{0.24\textwidth}\includegraphics[width=\textwidth]{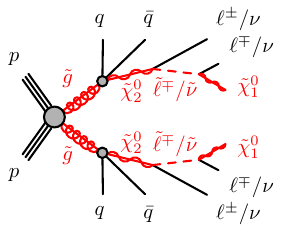}\caption{}\label{fig:feynman_gg2sl}\end{subfigure}  \\
\begin{subfigure}[t]{0.24\textwidth}\includegraphics[width=\textwidth]{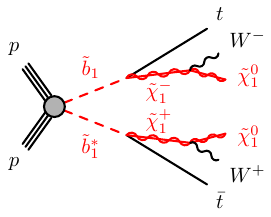}\caption{}\label{fig:feynman_b1b1}\end{subfigure} &
\begin{subfigure}[t]{0.24\textwidth}\includegraphics[width=\textwidth]{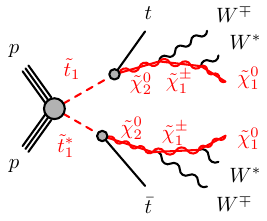}\caption{}\label{fig:feynman_t1t1}\end{subfigure} & 
\begin{subfigure}[t]{0.24\textwidth}\includegraphics[width=\textwidth]{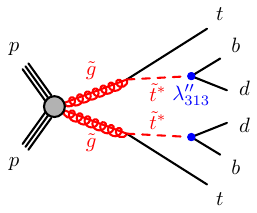}\caption{}\label{fig:feynm_rpv_gl313}\end{subfigure} &
\begin{subfigure}[t]{0.24\textwidth}\includegraphics[width=\textwidth]{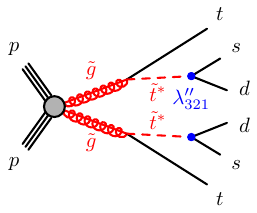}\caption{}\label{fig:feynm_rpv_gl321}\end{subfigure} \\
\begin{subfigure}[t]{0.24\textwidth}\includegraphics[width=\textwidth]{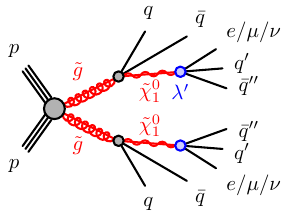}\caption{}\label{fig:feynm_rpv_glprime}\end{subfigure} & 
\begin{subfigure}[t]{0.24\textwidth}\includegraphics[width=\textwidth]{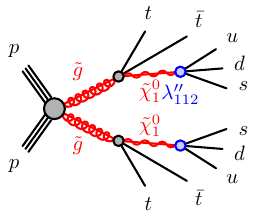}\caption{}\label{fig:feynm_rpv_gl112}\end{subfigure} &
\begin{subfigure}[t]{0.24\textwidth}\includegraphics[width=\textwidth]{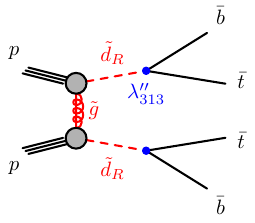}\caption{}\label{fig:feynm_rpv_sd313}\end{subfigure} &
\begin{subfigure}[t]{0.24\textwidth}\includegraphics[width=\textwidth]{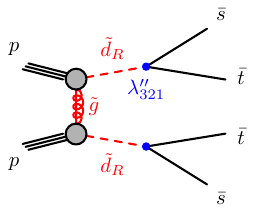}\caption{}\label{fig:feynm_rpv_sd321}\end{subfigure} \\
\end{tabular}
\caption{RPC SUSY processes featuring gluino ((a), (b), (c), (d)) or third-generation squark ((e), (f)) pair production studied in this analysis. 
RPV SUSY models considered are gluino pair production ((g), (h), (i), (j)) and t-channel production of down squark-rights ((k), (l)) 
which decay via baryon- or lepton-number violating couplings $\lambda''$ and $\lambda'$ respectively. In the diagrams, $q \equiv u, d, c, s$ 
and $\ell \equiv e,\mu,\tau$. In Figure~\ref{fig:feynman_gg2sl}, $\tilde{\ell} \equiv \tilde{e}, \tilde{\mu}, \tilde{\tau}$ and 
$\tilde{\nu} \equiv \tilde{\nu}_e, \tilde{\nu}_{\mu}, \tilde{\nu}_{\tau}$. In Figure~\ref{fig:feynman_t1t1}, the $W^*$ labels indicate 
largely off-shell $W$ bosons -- the mass difference between $\chinoonepm$ and $\ninoone$ is around 1~GeV.}
\label{fig:feynman}
\end{figure}

For RPC models, the first four scenarios studied focus on gluino pair production with decays into on-shell (Figure~\ref{fig:feynman_gtt}) 
or off-shell (Figure~\ref{fig:feynman_gttOffshell}) top quarks, as well as on-shell light quarks. The latter are accompanied by a cascade 
decay involving a $\chinoonepm$ and a $\ninotwo$ (Figure~\ref{fig:feynman_gg2WZ}) or a $\ninotwo$ and light sleptons (Figure~\ref{fig:feynman_gg2sl}). 
The other two RPC scenarios target the direct production of third-generation squark pairs 
with subsequent electroweakino-mediated decays (Figures~\ref{fig:feynman_b1b1} and~\ref{fig:feynman_t1t1}). 
The former is characterized by final states with bottom squark pairs decaying to $\ttbar WW\ninoone\ninoone$. The latter, addressed here by 
looking at a final state with three same-sign leptons, is a model that could explain the slight excess seen in same-sign lepton signatures 
during Run~1~\cite{Huang:2015fba}.
Finally, a full SUSY model with low fine-tuning, the non-universal Higgs model with two extra parameters (NUHM2)~\cite{Ellis:2002iu,Ellis:2002wv}, 
is also considered. When the soft-SUSY-breaking electroweakino mass, $m_{1/2}$, is in the range 300--800 GeV, the model mainly 
involves gluino pair production with gluinos decaying predominantly 
to $\ttbar\ninoone$ and $tb\chinoonepm$, giving rise to final states with two same-sign leptons and \met.

In the case of non-zero RPV couplings in the baryonic sector ($\lambda''_{ijk}$),
as proposed in scenarios with minimal flavour violation~\cite{Nikolidakis:2007fc,Smith:2008ju,Csaki:2011ge},
gluinos and squarks may decay directly to top quarks, leading to final states with same-sign leptons~\cite{Durieux:2013uqa,Berger:2013sir} 
and $b$-quarks (Figures~\ref{fig:feynm_rpv_gl313} and~\ref{fig:feynm_rpv_gl321}). Although these figures illustrate decay modes mediated by 
non-zero $\lambda''_{313}$ (resp. $\lambda''_{321}$) couplings, the exclusion limits set for these scenarios also hold for non-zero 
$\lambda''_{323}$ (resp. $\lambda''_{311}$ or $\lambda''_{322}$), as these couplings lead to experimentally indistinguishable final states.
Alternatively a gluino decaying to a neutralino LSP that further decays to SM particles via a non-zero RPV coupling in the leptonic sector, 
$\lambda'$, or in the baryonic sector $\lambda''$, is also possible (Figures~\ref{fig:feynm_rpv_glprime} and~\ref{fig:feynm_rpv_gl112}).  
Lower $\met$ is expected in these scenarios, as there is no stable LSP, and the \met originates from neutrinos produced 
in the $\ninoone$ and top quark decays. Pair production of  same-sign down squark-rights\footnote{These RPV baryon-number-violating couplings 
only apply to $SU(2)$ singlets.} 
(Figures~\ref{fig:feynm_rpv_sd313} and~\ref{fig:feynm_rpv_sd321}) is also considered.
In all of these scenarios, antisquarks decay into the charge-conjugate final states of those indicated for the corresponding squarks, 
and gluinos decay with equal probabilities into the given final state or its charge conjugate.

\section{ATLAS detector}
\label{sec:detector}

The ATLAS experiment~\cite{PERF-2007-01} is a multipurpose particle detector with a forward-backward symmetric cylindrical
geometry and nearly $4\pi$ coverage in solid angle.\footnote{ATLAS uses
  a right-handed coordinate system with its origin at the nominal
  interaction point (IP) in the centre of the detector and the
  $z$-axis along the beam pipe. The $x$-axis points from the IP to the
  centre of the LHC ring, and the $y$-axis points upward. Cylindrical
  coordinates ($r$, $\phi$) are used in the transverse plane, $\phi$
  being the azimuthal angle around the beam pipe. The pseudorapidity
  is defined in terms of the polar angle $\theta$ as $\eta = -\ln
  \tan(\theta/2)$. Rapidity is defined as $y=0.5 \ln\left[(E + p_z )/(E - p_z )\right]$ 
  where $E$ denotes the energy and $p_z$ is the component of the momentum along the beam direction. 
  The transverse momentum \pt, the transverse energy \et and the missing transverse momentum \met 
  are defined in the $x$--$y$ plane.}
The interaction point is surrounded by an inner detector (ID) for tracking, a
calorimeter system, and a muon spectrometer (MS).
The ID provides precision tracking of charged particles with
pseudorapidities $|\eta| < 2.5$ and is surrounded by a superconducting solenoid providing a \SI{2}{T} axial magnetic field.
It consists of silicon pixel and silicon micro-strip detectors inside a
transition radiation tracker. One significant upgrade for the $\sqrt{s}=13$~TeV running period is the presence of the
insertable B-Layer~\cite{ATLAS-TDR-19}, an additional pixel layer close to the interaction point, which 
provides high-resolution hits at small radius to improve the tracking and vertexing performance.
In the pseudorapidity region $|\eta| < 2.5$, high-granularity lead/liquid-argon 
electromagnetic sampling calorimeters are used.
A steel/scintillator tile calorimeter measures hadron energies for $|\eta| < 1.7$.
The endcap and forward regions, spanning $1.5<|\eta| <4.9$, are 
instrumented with liquid-argon calorimeters 
for both the electromagnetic and hadronic measurements. 
The MS consists of three large superconducting toroids
with eight coils each and
a system of trigger and precision-tracking chambers, 
which provide triggering and tracking capabilities in the
ranges $|\eta| < 2.4$ and $|\eta| < 2.7$, respectively.
A two-level trigger system is used to select events~\cite{Aaboud:2016leb}. The first-level
trigger is implemented in hardware. This is followed by the software-based high-level trigger,
which can run algorithms similar to those used in the offline reconstruction software, reducing the event rate to about \SI{1}{kHz}.

\section{Data set and simulated event samples}
\label{sec:dataMC}

The data used in this analysis were collected during 2015 and 2016 with a peak 
instantaneous luminosity of $L=1.4\times~10^{34}$~cm$^{-2}$s$^{-1}$. 
The mean number of $pp$ interactions per bunch crossing 
(pile-up) in the data set is 24. After the application of beam, detector and data-quality requirements, 
the integrated luminosity considered corresponds to 36.1~fb$^{-1}$.
The uncertainty in the combined 2015+2016 integrated luminosity is 3.2\%. 
It is derived, following a methodology similar to that detailed in Ref.~\cite{Aaboud:2016hhf}, 
from a preliminary calibration of the luminosity scale using $x$--$y$ beam-separation scans performed in August 2015 and May 2016.

Monte Carlo (MC) simulated event samples are used to model the SUSY signals and to estimate the irreducible SM background with two 
same-sign and/or three ``prompt'' leptons. Prompt leptons are produced directly in the hard-scattering process, or in the subsequent decays 
of $W$, $Z$ and $H$ bosons or prompt $\tau$ leptons. The reducible background, mainly 
arising from $\ttbar$ production, is estimated from the data as described in Section~\ref{sec:DD_bkg}. 
The MC samples were processed through a detailed ATLAS detector simulation~\cite{Aad:2010ah} based on 
\textsc{Geant4}~\cite{Agostinelli:2002hh} or a fast simulation using a parameterization of the calorimeter response 
and \textsc{Geant4} for the ID and MS~\cite{ATL-PHYS-PUB-2010-013}. To simulate the effects of additional $pp$ collisions 
in the same and nearby bunch crossings, inelastic interactions were generated using the soft strong-interaction processes 
of \PYTHIA 8.186~\cite{Sjostrand:2007gs} with a set of tuned parameters referred to as the A2 tune~\cite{ATL-PHYS-PUB-2012-003} and the MSTW2008LO parton 
distribution function (PDF) set~\cite{Martin:2009iq}. These MC events were overlaid onto the simulated hard-scatter event 
and reweighted to match the pile-up conditions observed in the data. 
Table~\ref{tab:MC} presents, for all samples, the event generator, parton shower, cross-section normalization, PDF
set and the set of tuned parameters for the modelling of the parton shower, hadronization and underlying event. 
In all MC samples, except those produced by the \SHERPA event generator, 
the \textsc{EvtGen}~v1.2.0 program~\cite{EvtGen} was used 
to model the properties of bottom and charm hadron decays. 

\begin{table*}[ht]
\begin{center}
\scriptsize
\resizebox{\textwidth}{!}
{
\begin{tabular}{|l|l|c|c|c|c|}
\hline
Physics process    & Event generator & Parton shower & Cross-section & PDF set & Set of tuned \\
                   &	      & 	      & normalization & 	& parameters  \\
\hline
\hline
Signal            &                      			&                      			&                               	&               &      \\
RPC   	   	& \AMCATNLO 2.2.3~\cite{Alwall:2014hca} 	& \PYTHIA 8.186~\cite{Sjostrand:2007gs}	& NLO+NLL  & NNPDF2.3LO~\cite{Ball:2012cx} & A14~\cite{ATL-PHYS-PUB-2014-021} \\
RPV except Fig.~\ref{fig:feynm_rpv_gl112} & \AMCATNLO 2.2.3     & \PYTHIA 8.210       			& or			 		& NNPDF2.3LO	& A14 \\
RPV Fig.~\ref{fig:feynm_rpv_gl112}   & \textsc{Herwig++} 2.7.1~\cite{Corcella:2000bw}       & \textsc{Herwig++} 2.7.1  	& NLO-Prospino2 ~\cite{Beenakker:1996ch,Kulesza:2008jb,Kulesza:2009kq,Beenakker:2009ha,Beenakker:2011fu,Kramer:2012bx}     				&CTEQ6L1~\cite{Pumplin:2002vw} & UEEE5~\cite{Gieseke:2012ft}  \\
\hline
\hline
$\ttbar +X$            &                      			&                      			&                               	&               &      \\
$\ttbar W$, $\ttbar Z/\gamma^{*}$ & \AMCATNLO 2.2.2 		& \PYTHIA 8.186	  			& NLO~\cite{YR4}     			& NNPDF2.3LO	& A14    \\
$\ttbar H$	   & \AMCATNLO 2.3.2        			& \PYTHIA 8.186  			& NLO~\cite{YR4}   			& NNPDF2.3LO	& A14  \\
4$t$    	& \AMCATNLO 2.2.2       			& \PYTHIA 8.186        			& NLO~\cite{Alwall:2014hca}	  	& NNPDF2.3LO	& A14  \\
\hline
Diboson            &                      			&                      			&                               	&               &      \\
$ZZ$, $WZ$    	   & \SHERPA 2.2.1~\cite{gleisberg:2008ta}      & \SHERPA 2.2.1				& NLO~\cite{ATL-PHYS-PUB-2016-002}	&NNPDF2.3LO & \SHERPA default \\
Other (inc. $W^{\pm}W^{\pm}$)   & \SHERPA 2.1.1 		& \SHERPA 2.1.1				& NLO~\cite{ATL-PHYS-PUB-2016-002}	&CT10~\cite{Lai:2010vv} & \SHERPA default \\
\hline
Rare               &                      			&                      			&                               	&               &      \\
$\ttbar WW$, $\ttbar WZ$     & \AMCATNLO 2.2.2       		& \PYTHIA 8.186      			& NLO~\cite{Alwall:2014hca}	  	& NNPDF2.3LO	 & A14  \\
$tZ$, $tWZ$, $t\ttbar$    & \AMCATNLO 2.2.2        		& \PYTHIA 8.186       			& LO		                   	& NNPDF2.3LO     & A14  \\
$WH$, $ZH$	   & \AMCATNLO 2.2.2        			& \PYTHIA 8.186      			& NLO~\cite{Dittmaier:2012vm}   	& NNPDF2.3LO     & A14  \\
Triboson	   & \SHERPA 2.1.1         			& \SHERPA 2.1.1        			& NLO~\cite{ATL-PHYS-PUB-2016-002}       & CT10	     	& \SHERPA default \\
\hline
\end{tabular}
}
\caption{Simulated signal and background event samples: the corresponding event generator, parton shower, cross-section normalization, PDF set and 
set of tuned parameters are shown for each sample. Because of their very small contribution to the signal-region background estimate, 
$\ttbar WW$, $\ttbar WZ$, $tZ$, $tWZ$, $t\ttbar$, $WH$, $ZH$ and triboson are summed and labelled ``rare'' in the following. 
NLO-Prospino2 refers to RPV down squark models of Figures~\ref{fig:feynm_rpv_sd313} and \ref{fig:feynm_rpv_sd321}, as well as the NUHM2 model.}
\label{tab:MC}
\end{center}
\end{table*}

The SUSY signals from Figure~\ref{fig:feynman} are defined by an effective Lagrangian describing the interactions of a small number of new 
particles~\cite{Alwall:2008ve,Alwall:2008ag,Alves:2011wf}. All SUSY particles not included 
in the decay of the pair-produced squarks and gluinos are effectively decoupled. These simplified models assume one 
production process and one decay channel with a 100\% branching fraction. Apart from Figure~\ref{fig:feynm_rpv_gl112}, where 
events were generated with \textsc{Herwig++}~\cite{Corcella:2000bw}, all simplified models were generated from leading-order (LO) matrix elements 
with up to two extra partons in the matrix element (only up to one for the $\gluino\to q\bar q(\ell\ell/\nu\nu)\ninoone$ model) 
using \AMCATNLO 2.2.3~\cite{Alwall:2014hca} interfaced to \PYTHIA 8 with the A14 tune~\cite{ATL-PHYS-PUB-2014-021} for the 
modelling of the parton shower, hadronization and underlying event.
Jet--parton matching was realized following the CKKW-L prescription~\cite{Lonnblad:2011xx}, with a matching scale set to one quarter of 
the pair-produced superpartner mass. All signal models were generated 
with prompt decays of the SUSY particles. Signal cross-sections were calculated at next-to-leading order (NLO) in the strong coupling constant, 
adding the resummation of soft-gluon emission at next-to-leading-logarithmic 
accuracy (NLO+NLL)~\cite{Beenakker:1996ch,Kulesza:2008jb,Kulesza:2009kq,Beenakker:2009ha,Beenakker:2011fu}, except 
for the RPV models of Figures~\ref{fig:feynm_rpv_sd313} and~\ref{fig:feynm_rpv_sd321} and the NUHM2 model where NLO 
cross-sections were used~\cite{Beenakker:1996ed,Beenakker:1996ch}. The nominal cross-sections and the uncertainties were taken from 
envelopes of cross-section predictions using different PDF sets 
and factorization and renormalization scales, as described in Refs.~\cite{Kramer:2012bx,Borschensky:2014cia}. 
Typical pair-production cross-sections are: $4.7 \pm 1.2$~fb for gluinos with a mass of \SI{1.7}{\TeV}, $28 \pm 4$~fb
for bottom squarks with a mass of \SI{800}{\GeV}, and $15.0\pm 2.0$~fb for down 
squark-rights with a mass of \SI{800}{\GeV} and a gluino mass of \SI{2.0}{\TeV}.

The two dominant irreducible background processes are $\ttbar V$ (with $V$ being a $W$ or $Z/\gamma^*$ boson) 
and diboson production with final states of four charged leptons $\ell$,\footnote{All lepton flavours are included here and $\tau$
leptons subsequently decay leptonically or hadronically.} three charged leptons and one neutrino, or 
two same-sign charged leptons and two neutrinos. The MC simulation samples for these are described in 
Refs.~\cite{ATL-PHYS-PUB-2016-005} and~\cite{ATL-PHYS-PUB-2016-002}, 
respectively. For diboson production, the matrix elements contain the doubly resonant diboson processes and all other diagrams with four 
or six electroweak vertices, such as $W^\pm W^\pm jj$, with one ($W^\pm W^\pm jj$) or two ($WZ$, $ZZ$) extra partons.
NLO cross-sections for $\ttbar W$, $\ttbar Z/\gamma^*(\rightarrow \ell \ell)$\footnote{This cross-section is computed 
using the configuration described in Refs.~\cite{Alwall:2014hca,Frixione:2015zaa}.} 
and leptonic diboson processes are respectively 0.60~pb~\cite{YR4}, 0.12~pb and 
6.0~pb~\cite{ATL-PHYS-PUB-2016-002}. The processes $\ttbar H$ and 4$t$, with NLO cross-sections of 507.1~fb~\cite{YR4} and 
9.2~fb~\cite{Alwall:2014hca} respectively, are also considered.

Other background processes, with small cross-sections and no significant contribution to any of the signal regions, 
are grouped into a category labelled ``rare''. This category contains 
$\ttbar WW$ and $\ttbar WZ$ events generated with no extra parton in the matrix element, and $tZ$, $tWZ$, $t\ttbar$, $WH$ and $ZH$ as well as 
triboson ($WWW$, $WWZ$, $WZZ$ and $ZZZ$) production with fully leptonic decays, leading to up to six charged leptons. 
The processes $WWW$, $WZZ$ and $ZZZ$ were generated at NLO 
with additional LO matrix elements for up to two extra partons, 
while $WWZ$ was generated at LO with up to two extra partons.

\section{Event reconstruction and selection}

\label{sec:selection}

Candidate events are required to have a reconstructed vertex~\cite{ATL-PHYS-PUB-2015-026} 
with at least two associated tracks with $\pt >400$~MeV. The vertex with the largest $\Sigma \pt^2$ of the associated tracks 
is chosen as the primary vertex of the event.

For the data-driven background estimations, two categories of electrons and muons are used: 
``candidate'' and ``signal'' with the latter being a subset of the ``candidate'' leptons satisfying tighter selection criteria. 
Electron candidates are reconstructed from energy depositions in the electromagnetic calorimeter which were matched to an ID track 
and are required to have $|\eta|<2.47$, $\pT>\SI{10}{\GeV}$,
and pass the ``Loose'' likelihood-based identification requirement~\cite{ATLAS-CONF-2016-024}. 
Candidates within the transition region between the barrel and endcap electromagnetic calorimeters,
$1.37<|\eta|<1.52$, are not considered. The track matched with the electron must have a significance of the transverse impact parameter $d_0$ 
with respect to the reconstructed primary vertex of $\vert d_0\vert/\sigma(d_0) < 5$.
Muon candidates are reconstructed in the region $|\eta|<2.5$ 
from muon spectrometer tracks matching ID tracks.
All muon candidates must have $\pT>\SI{10}{\GeV}$ and must pass the ``Medium'' identification requirements~\cite{Aad:2016jkr}.

Jets are reconstructed with the anti-$k_t$
algorithm~\cite{Cacciari:2008} with radius parameter $R=0.4$, using three-dimensional topological energy
clusters in the calorimeter~\cite{PERF-2014-07} as input. 
All jets must have $\pT>\SI{20}{\GeV}$ and $|\eta|<2.8$.
For all jets the expected average energy contribution from
pile-up is subtracted according to the jet area~\cite{Cacciari:2007fd,Aaboud:2017jcu}.
Jets are then calibrated as described in Ref.~\cite{Aaboud:2017jcu}. In order to reduce the effects of pile-up, 
a significant fraction of the tracks in jets with $\pt<\SI{60}{GeV}$ and $|\eta|<2.4$ must originate from the primary vertex, 
as defined by the jet vertex tagger (JVT)~\cite{ATLAS-CONF-2014-018}. 

Identification of jets containing $b$-hadrons ($b$-tagging) is performed with the MV2c10 algorithm, 
a multivariate discriminant making use of track impact parameters 
and reconstructed secondary vertices~\cite{Aad:2015ydr,ATL-PHYS-PUB-2015-022}.
A requirement is chosen corresponding to a 70\% average efficiency 
for tagging $b$-jets in simulated $\ttbar$ events. 
The rejection factors for light-quark/gluon jets, $c$-quark jets and $\tau \ra \nu+\textrm{hadron}$ decays in simulated $\ttbar$ events 
are approximately 380, 12 and 54, respectively~\cite{ATL-PHYS-PUB-2015-022,ATL-PHYS-PUB-2016-012}. 
Jets with $|\eta|<2.5$ which satisfy the $b$-tagging and JVT requirements are identified as $b$-jets. 
Correction factors and uncertainties determined from data for the $b$-tagging efficiencies and mis-tag rates
are applied to the simulated samples~\cite{ATL-PHYS-PUB-2015-022}. 

After the object identification, overlaps between the different objects are resolved. 
Any jet within a distance $\Delta R_y \equiv \sqrt{(\Delta y)^2+(\Delta\phi)^2} =$ 0.2 of a lepton candidate is discarded, 
unless the jet is $b$-tagged,\footnote{In this case the $b$-tagging operating point corresponding to an efficiency of 85\% is used.} 
in which case the lepton is discarded since it probably originated from a semileptonic $b$-hadron decay. 
Any remaining lepton within $\Delta R_y = \operatorname{min}\{0.4, 0.1 + \SI{9.6}{GeV}/\pt(\ell)\}$ of a jet is discarded. 
In the case of muons, the muon is retained and the jet is discarded if the jet has fewer than three associated tracks. This reduces 
inefficiencies for high-energy muons undergoing significant energy loss in the calorimeter. 

Signal electrons must satisfy the ``Medium'' likelihood-based identification requirement~\cite{ATLAS-CONF-2016-024}.
In regions with large amounts of material in the tracker, an electron (positron) is more likely to emit a hard brems\-strah\-lung photon; 
if the photon subsequently converts to an asymmetric electron--positron pair, 
and the positron (electron) has high momentum and is reconstructed, the lepton charge can be misidentified 
(later referred to as ``charge-flip''). 
To reduce the impact of charge misidentification, signal electrons must satisfy $|\eta|<2.0$. Furthermore, signal electrons that 
are likely to be reconstructed with an incorrect charge assignment are rejected using the electron
cluster and track properties including the impact parameter, the curvature significance, the cluster width, and the
quality of the matching between the cluster and its associated track, in terms of both energy and position. These
variables, as well as the electron \pt and $\eta$, are combined into a single classifier using a boosted decision tree (BDT) algorithm. 
A selection requirement on the BDT output is chosen to achieve a rejection factor of 7--8 for electrons 
with a wrong charge assignment while selecting correctly measured electrons with an efficiency of 97\%. Correction
factors to account for differences in the selection efficiency between data and MC simulation are applied to the selected 
electrons in MC simulation. These correction factors are determined using $Z\ra ee$ events~\cite{Aaboud:2016vfy}.

Signal muons must fulfil the requirement $\vert d_0\vert/\sigma(d_0) < 3$. 
Tracks associated with the signal electrons or muons must have a longitudinal impact parameter $z_0$ with respect to the 
reconstructed primary vertex satisfying $\vert z_0 \sin\theta\vert  < 0.5$~mm. 
Isolation requirements are applied to both the signal electrons and muons. 
The  scalar sum of the \pt of tracks within a variable-size cone around the lepton, 
excluding its own track, must be less than 6\% of the lepton \pt. 

The track isolation cone size for electrons (muons) 
$\Delta R_\eta \equiv \sqrt{(\Delta\eta)^2+(\Delta\phi)^2}$ 
is given by the smaller of $\Delta R_\eta = \SI{10}{~GeV}/\pt$ and $\Delta R_\eta = 0.2\,(0.3)$. 
In addition, in the case of electrons the calorimeter energy clusters in a cone of $\Delta R_\eta = 0.2$ around the electron 
(excluding the deposit from the electron itself) must be less than 6\% of the electron \pt. 
Simulated events are corrected to account for differences in the lepton trigger, reconstruction, 
identification and isolation efficiencies between data and MC simulation.

The missing transverse momentum is defined as the negative vector sum of the transverse momenta 
of all identified candidate objects (electrons, photons~\cite{Aaboud:2016yuq}, muons and jets) and an additional soft term. 
The soft term is constructed from all tracks associated with the primary vertex but not with any physics object. 
In this way, the $\met$ is adjusted for the best calibration of the jets and the other identified physics objects listed above, 
while maintaining approximate pile-up independence in the soft term~\cite{ATL-PHYS-PUB-2015-027, ATL-PHYS-PUB-2015-023}.

Events are selected using a combination of dilepton and $\met$ triggers, the latter being used only for events with $\met>\SI{250}{GeV}$. 
The trigger-level requirements on $\met$ and the leading and subleading lepton \pt are looser than those applied offline 
to ensure that trigger efficiencies are constant in the relevant phase space. The event selection requires at least two signal 
leptons with $\pt>20$~\GeV~(apart from two signal regions where the lower bound on the subleading lepton \pt is 10~\GeV).\footnote{To ensure 
that the trigger efficiency is constant for selected events where the subleading lepton \pt lies between 10 and 20 GeV only 
the \met trigger is used in this case.} If the event contains exactly two signal leptons, they must have the same electric charge. 
In order to reject detector noise and non-collision backgrounds (including those from cosmic rays, beam-gas and beam-halo interactions), 
events are discarded if they contain any jet not satisfying basic quality criteria~\cite{DAPR-2012-01, PERF-2012-01}.

To maximize the sensitivity to the signal models of Figure~\ref{fig:feynman}, 19 non-exclusive\footnote{Each signal 
region partially overlaps with at least one other signal region.} signal regions (SRs) are defined in Table~\ref{tab:SRdef3}. 
The SRs are named in the form \textit{SN}L{\textit M}b{\textit X}, where {\textit S} indicates if the signal region is targeting 
an RPC or RPV model, {\textit N} indicates the number of leptons required, {\textit M} the number of $b$-jets required, and {\textit X} indicates the severity 
of the \met or \meff\ requirements (Soft, Medium or Hard). All signal regions, except Rpv2L0b, allow any number of additional leptons in addition to a $e^\pm e^\pm$,
$e^\pm \mu^\pm$ or $\mu^\pm \mu^\pm$ pair. Signal regions with a three lepton selection can either require any lepton charge combination (Rpc3L0bH, Rpc3L0bS) or 
that all three leptons have the same charge (Rpc3LSS1b). The other requirements used to define the SRs are the number of signal leptons 
($N_{\textrm{leptons}}^{\textrm{signal}}$), number of $b$-jets with $\pt>\SI{20}{\GeV}$ ($N_{b\textrm{-jets}}$), number of jets with \pt above 25, 40 
or \SI{50}{GeV}, regardless of their flavour ($N_{\textrm{jets}}$), \met, the effective mass (\meff) and the charge of the signal leptons. 
The \meff\ variable is defined as the scalar sum of the $\pt$ of the signal leptons, jets and the \met. 
For SRs where the $Z$+jets background is important (Rpc3LSS1b, Rpv2L0b and Rpv2L2bH), 
events in which the invariant mass of two same-sign electrons is close to the $Z$ boson mass are vetoed. 
For SRs targeting the production of down squark pairs (Rpv2L1bS, Rpv2L2bS, Rpv2L1bM),
only events with at least two negatively charged leptons are considered, as the down squarks decay exclusively to top antiquarks.
Finally, SRs targeting signal scenarios with lepton \pt spectra softer than typical background processes
impose an upper bound on the leptons' \pt.  The last column of Table~\ref{tab:SRdef3} indicates the targeted signal model. The Rpc3L1b and Rpc3L1bH SRs 
are not motivated by a particular signal model and can be seen as a natural extension of the Rpc3L0b SRs with the same kinematic selections 
but requiring at least one $b$-jet.

The values of acceptance times efficiency of the SR selections for the RPC SUSY signal models, with masses near the exclusion limit, 
typically range between 0.5\% and 7\% for models with a light \ninoone and between 0.5 and 2\% for models with a heavy \ninoone. For RPV SUSY signal models, 
these values are in the range 0.2--4\%. To increase the signal efficiency for the SUSY models with low-energy leptons 
(Figure~\ref{fig:feynman_gttOffshell}), the \pt threshold of leptons is relaxed from 20~GeV to 10~GeV in the SR definition. 

\begin{table}[tbh!]
\centering
\resizebox{\textwidth}{!}
{
\hspace{0.5cm}
\def\arraystretch{1.2}
\small
\begin{tabular}{|l|c|c|c|c|c|r|c|c|l|}
\hline
Signal region  &  $N_{\textrm{leptons}}^{\textrm{signal}}$   & $N_{b\textrm{-jets}}$ & $N_{\textrm{jets}}$  & $\pt^{\textrm{jet}}$ & \met\ & \meff\ & \met/\meff  & Other & Targeted  \\
               &                                  &                   &                  &    [GeV]             & [GeV] & [GeV] &   &  & Signal  \\
\hline\hline

Rpc2L2bS         & $\ge 2$SS  & $\ge 2$ & $\ge 6$ & $>25$ & $>200$ & $>600$  & $>0.25$    & --			        & Fig.~\ref{fig:feynman_gtt}\\ 
Rpc2L2bH         & $\ge 2$SS  & $\ge 2$ & $\ge 6$ & $>25$ & --     & $>1800$  & $>0.15$	  & -- 			        & Fig.~\ref{fig:feynman_gtt}, NUHM2\\ 
\hline
Rpc2Lsoft1b    & $\ge 2$SS  & $\ge 1$ & $\ge 6$ & $>25$ & $>100$ &  --\hphantom{00}      & $>0.3\hphantom{0}$    & 20,10 $<$\ptlone,\ptltwo $<$ 100 GeV & Fig.~\ref{fig:feynman_gttOffshell}\\ 
Rpc2Lsoft2b      & $\ge 2$SS  & $\ge 2$ & $\ge 6$ & $>25$ & $>200$ & $>600$   & $>0.25$   & 20,10 $<$\ptlone,\ptltwo $<$ 100 GeV & Fig.~\ref{fig:feynman_gttOffshell} \\ 
\hline
Rpc2L0bS         & $\ge 2$SS  & $=0$    & $\ge 6$ & $>25$ & $>150$ & --\hphantom{00}      & $>0.25$   & -- 				& Fig.~\ref{fig:feynman_gg2WZ}\\
Rpc2L0bH         & $\ge 2$SS  & $=0$    & $\ge 6$ & $>40$ & $>250$ & $>900$   & --	  & --				& Fig.~\ref{fig:feynman_gg2WZ}\\
\hline
Rpc3L0bS       & $\ge 3$    & $=0$    & $\ge 4$ & $>40$ & $>200$ & $>600$   & --	  & --				& Fig.~\ref{fig:feynman_gg2sl}\\ 
Rpc3L0bH       & $\ge 3$    & $=0$    & $\ge 4$ & $>40$ & $>200$ & $>1600$  & --  & --				& Fig.~\ref{fig:feynman_gg2sl}\\
Rpc3L1bS       & $\ge 3$    & $\ge 1$ & $\ge 4$ & $>40$ & $>200$ & $>600$   & --  & --				& Other \\ 
Rpc3L1bH       & $\ge 3$    & $\ge 1$ & $\ge 4$ & $>40$ & $>200$ & $>1600$  & --  & --				& Other  \\
\hline
Rpc2L1bS         & $\ge 2$SS  & $\ge 1$ & $\ge 6$ & $>25$ & $>150$ & $>600$   & $>0.25$   & --				& Fig.~\ref{fig:feynman_b1b1}\\
Rpc2L1bH         & $\ge 2$SS  & $\ge 1$ & $\ge 6$ & $>25$ & $>250$ & --\hphantom{00}      & $>0.2\hphantom{0}$    & --				& Fig.~\ref{fig:feynman_b1b1}\\ 
\hline
Rpc3LSS1b    & $\ge \ell^\pm\ell^\pm\ell^\pm$ & $\ge 1$ & -- & --   & --  & --\hphantom{00}       & -- & veto 81$<$\mee$<$101 GeV 	& Fig.~\ref{fig:feynman_t1t1}\\ 
\hline
Rpv2L1bH       & $\ge 2$SS  & $\ge 1$ & $\ge 6$ & $>50$ & --     & $>2200$  & --        &  --				& Figs.~\ref{fig:feynm_rpv_gl313}, \ref{fig:feynm_rpv_gl321}\\
Rpv2L0b        & $=2$SS     & $=0$    & $\ge 6$ & $>40$ & --     & $>1800$  & --        &  veto 81$<$\mee$<$101 GeV 		& Fig.~\ref{fig:feynm_rpv_glprime} \\
Rpv2L2bH       & $\ge 2$SS  & $\ge 2$ & $\ge 6$ & $>40$ & --     & $>2000$  & --      & veto 81$<$\mee$<$101 GeV		& Fig.~\ref{fig:feynm_rpv_gl112} \\
Rpv2L2bS       & $\ge \ell^-\ell^-$   & $\ge 2$ & $\ge 3$ & $>50$ & --     & $>1200$   & --                          & --      & Fig.~\ref{fig:feynm_rpv_sd313}\\
Rpv2L1bS   & $\ge \ell^-\ell^-$  & $\ge 1$ & $\ge 4$ & $>50$ & --     & $>1200$  & --        & -- 			        & Fig.~\ref{fig:feynm_rpv_sd321}\\
Rpv2L1bM  & $\ge \ell^-\ell^-$  & $\ge 1$ & $\ge 4$ & $>50$ & --     & $>1800$  & --        & --			        & Fig.~\ref{fig:feynm_rpv_sd321}\\
\hline
\end{tabular}
}
\caption{Summary of the signal region definitions. Unless explicitly stated in the table, at least two signal leptons with 
$\pt>$20 GeV and same charge (SS) are required in each signal region. Requirements 
are placed on the number of signal leptons ($N_{\textrm{leptons}}^{\textrm{signal}}$), the number of 
$b$-jets with $\pt>\SI{20}{GeV}$ ($N_{b\textrm{-jets}}$), the number of jets ($N_{\textrm{jets}}$) above a certain \pt threshold ($\pt^{\textrm{jet}}$), 
\met, \meff\ and/or \met/\meff. The last column indicates the targeted signal model. The Rpc3L1b and Rpc3L1bH SRs 
are not motivated by a particular signal model and can be seen as a natural extension of the Rpc3L0b SRs with the same kinematic selections 
but requiring at least one $b$-jet.}
\label{tab:SRdef3}
\end{table}

\section{Background estimation}
\label{sec:bkg}

Two main sources of SM background can be distinguished in this analysis. The first category is the reducible  
background, which includes events containing electrons with mismeasured charge, mainly from the production of top quark pairs, 
and events containing at least one fake or non-prompt (FNP) lepton. The FNP lepton mainly originates from heavy-flavour hadron decays 
in events containing top quarks, or $W$ or $Z$ bosons. Hadrons misidentified as leptons, electrons from photon conversions and leptons 
from pion or kaon decays in flight are other possible sources. Data-driven methods used for the estimation of this reducible background  in the signal 
and validation regions are described in Section~\ref{sec:DD_bkg}. 

The second background category is the irreducible background from events with two same-sign prompt 
leptons or at least three prompt leptons and is estimated using the MC simulation samples. Since diboson and $\ttbar V$ events are the main 
irreducible backgrounds in the signal regions, dedicated validation regions (VR) with an enhanced contribution from these processes, and small 
signal contamination, are defined to verify the background predictions from the simulation (Section~\ref{sec:valid}). 
Section~\ref{sec:syst} discusses the systematic uncertainties considered when performing the background estimation in the signal 
and validation regions.

\subsection{Reducible background estimation methods} 
\label{sec:DD_bkg}

Charge misidentification is only relevant for electrons. The contribution of charge-flip events 
to the SR/VR is estimated using the data. 
The electron charge-flip probability is extracted in a $Z/\gamma^{*}\to ee$ data sample using a likelihood fit 
which takes as input the numbers of same-sign and opposite-sign electron pairs observed in a window of 10~GeV around the $Z$ boson mass. 
The charge-flip probability is a free parameter of the fit and is extracted as a function of the electron $\pt$ and $\eta$. 
These probabilities are around 0.5\% (1\%) and 0.1\% (0.2\%) for the candidate and signal electrons for $|\eta|<1.37$ ($|\eta|>1.52$), respectively. 
The former is used only in the FNP lepton background estimation.
The event yield of the charge-flip electron background in the signal or validation regions is obtained by multiplying the 
measured charge-flip probability with the number of events in data regions with the same kinematic requirements 
as the signal or validation regions but with opposite-sign lepton pairs.

Two data-driven methods are used to estimate the FNP lepton background, referred to as the ``matrix method'' and the ``MC template method''. 
The estimates from these methods are combined to give the final estimate. These two methods are described below. 

The first estimation of the FNP lepton background is performed with a matrix method similar to that described in Ref.~\cite{SUSY3bjetsRun1}. 
Two types of lepton identification criteria are defined: ``tight'', 
corresponding to the signal lepton criteria described in Section~\ref{sec:selection}, 
and ``loose'', corresponding to candidate leptons after object overlap removal and the charge-flip BDT selection described also in Section~\ref{sec:selection}. 
The matrix method relates the number of events containing prompt or FNP leptons 
to the number of observed events with tight or loose-not-tight leptons 
using the probability for loose prompt or FNP leptons to satisfy the tight criteria. 
The probability for loose prompt leptons to satisfy the tight selection criteria ($\varepsilon$)
is obtained using a $Z/\gamma^*\to\ell\ell$ data sample and is modelled as a function of the lepton $\pt$ and $\eta$.
The efficiencies for electrons (muons) rise from 60\% (80\%) at low \pt to almost 100\% at \pt above 50 GeV -- apart from 
endcap electrons, for which they reach only 95\%. The probability for loose FNP leptons to satisfy the tight selection criteria 
(FNP lepton rate, $f$) is determined from data in SS control regions enriched in non-prompt leptons mostly originating 
from heavy-flavour hadron decays in single-lepton $\ttbar$ events. These regions contain events with at least one $b$-jet, 
one well-isolated muon (referred to as the ``tag''), and an additional loose electron or muon which is used for the measurement. 
The rates $f$ are measured as a function of \pt after subtracting the small contribution from prompt-lepton processes predicted by simulation 
and the data-driven estimation of events with electron charge-flip.\footnote{For muons with $\pt<20$~GeV, $f$ is parameterized as a 
function of $\pt$ and $\eta$.} For electrons, and muons with $|\eta|<2.3$, $f$ is constant at around 10\% for $\pt<30$~GeV (20\% for muons with
$|\eta|>2.3$) and increases at higher \pt. With these values of $\varepsilon$ and $f$, the method has been demonstrated to correctly estimate the FNP
lepton background.

The second method for FNP lepton estimation is the MC template method described in details in Refs.~\cite{ATLAS-CONF-2012-151,SUSY3bjetsRun1}. 
It relies on the correct modelling of the kinematic distributions of the FNP leptons and charge-flipped electron processes in $\ttbar$ 
and $V$+jets samples. These samples were simulated with the \POWHEGBOX generator~\cite{powheg,Frixione:2007vw,Alioli:2010xd,Campbell:2014kua} 
and the parton shower and hadronization performed by either \PYTHIA 6.428~\cite{Sjostrand:2006za} ($t\bar t$) or \PYTHIA 8.186 ($V$+jets).  
The FNP leptons are classified in five categories, namely electrons and muons originating 
from $b$- and light-quark jets as well as electrons from photon conversions. Normalization factors for each of the five sources are 
adjusted to match the observed data in dedicated control regions. Events are selected with at least two same-sign signal leptons,
$\met>40$~GeV, two or more jets, and are required not to belong to the SRs. They are further split into regions 
with or without $b$-jets and with different lepton flavours of the same-sign lepton pair, giving a total of six control regions. 
The global normalization factors applied to the MC samples for estimating the reducible background in each SR 
vary from $1.2\pm1.1$ to $2.9\pm2.0$, where the errors account for statistical uncertainties and uncertainties related to the choice 
of event generator (see Section~\ref{sec:syst}).

\begin{figure}[th!]
\centering
\begin{subfigure}[t]{0.49\textwidth}\includegraphics[width=\textwidth]{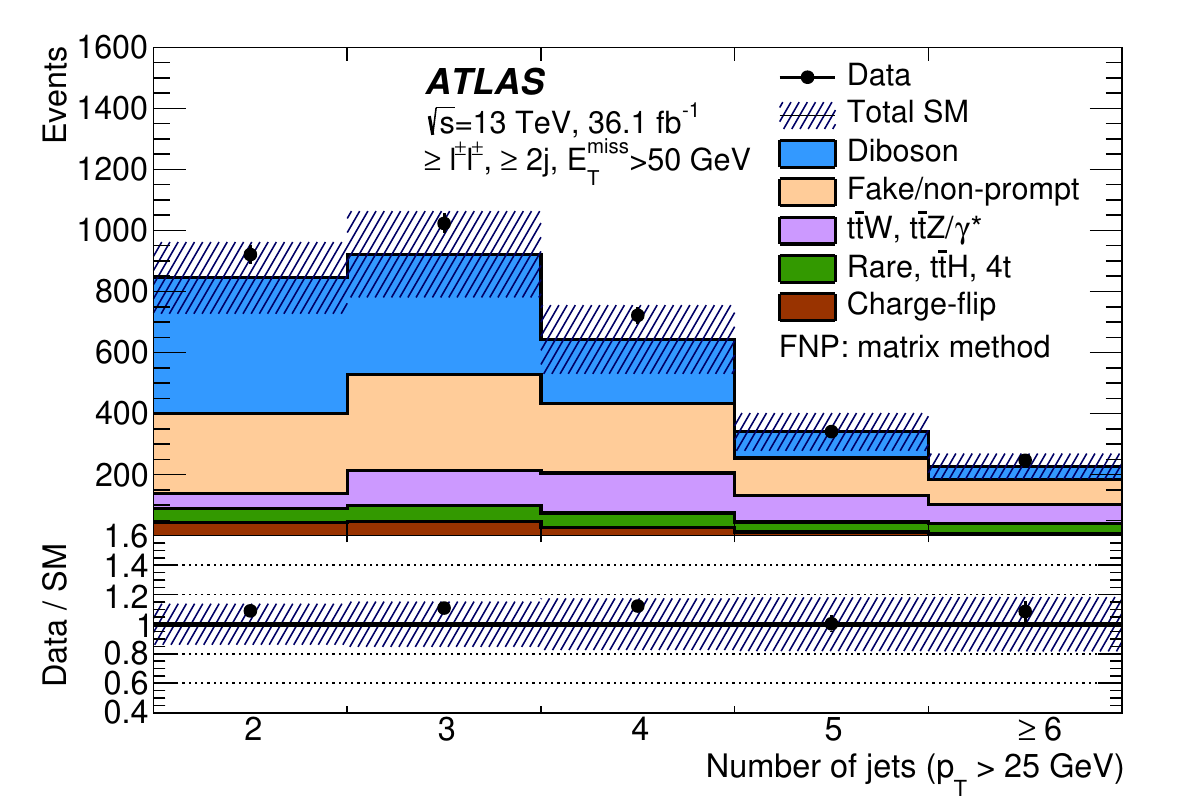}\caption{}\label{fig:VRnj}\end{subfigure}
\begin{subfigure}[t]{0.49\textwidth}\includegraphics[width=\textwidth]{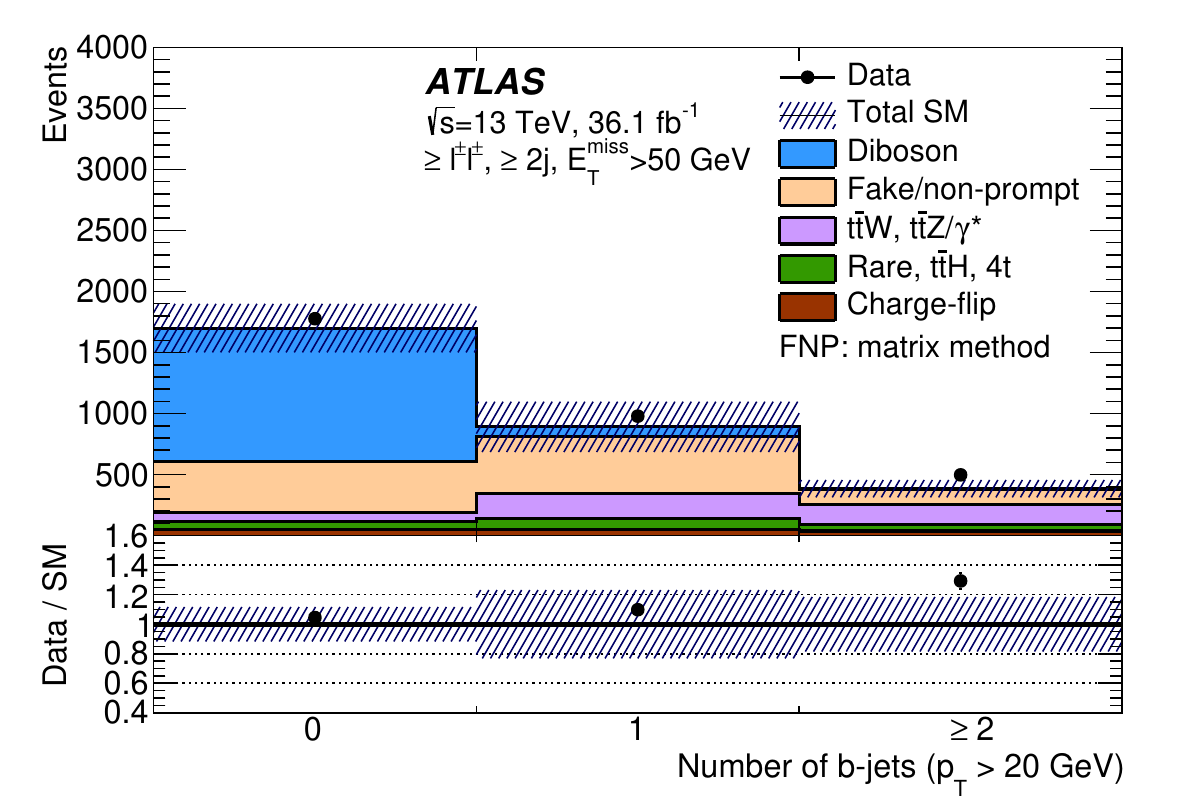}\caption{}\label{fig:VRnb}\end{subfigure}
\begin{subfigure}[t]{0.49\textwidth}\includegraphics[width=\textwidth]{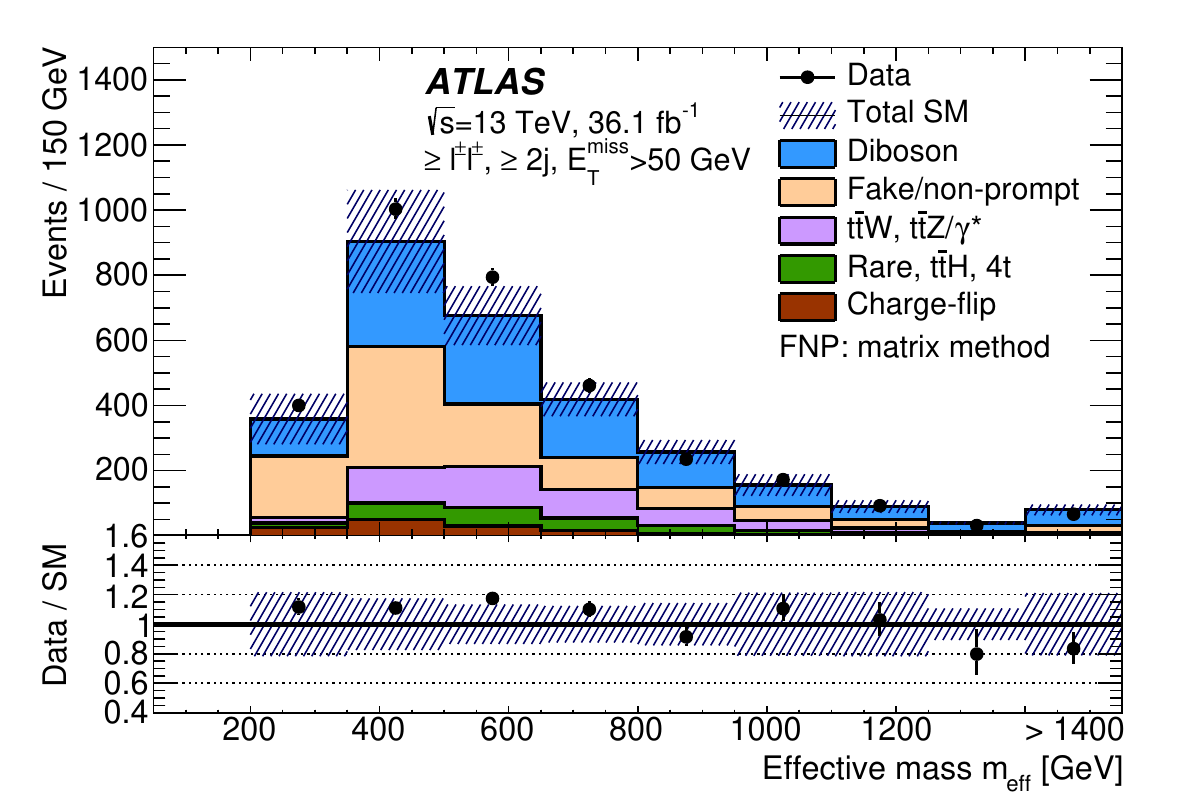}\caption{}\label{fig:VRmeff1}\end{subfigure}
\begin{subfigure}[t]{0.49\textwidth}\includegraphics[width=\textwidth]{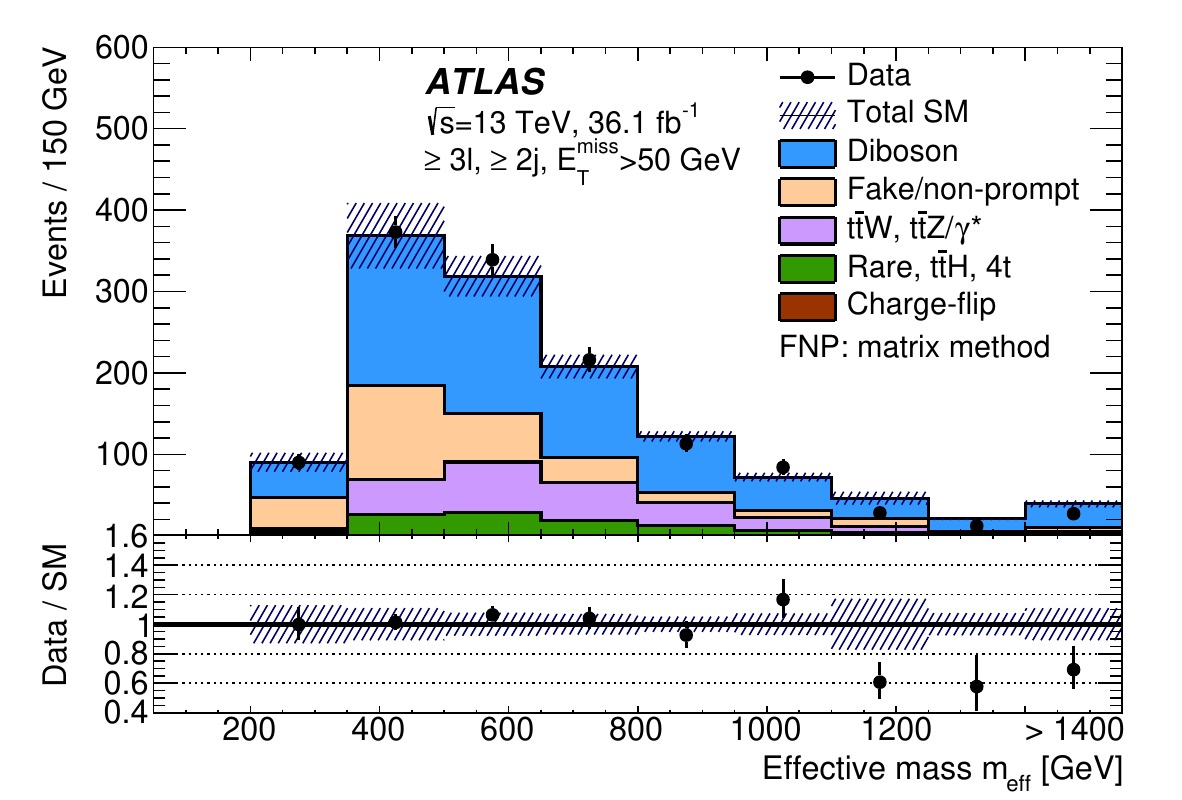}\caption{}\label{fig:VRmeff2}\end{subfigure}
\caption{
Distributions of (a) the number of jets, (b) the number of $b$-tagged jets and (c), (d) the effective mass. The distributions are made 
after requiring at least two jets ($\pT>\SI{40}{GeV}$) and $\met>\SI{50}{GeV}$, as well as at least two same-sign leptons (a, b, c) 
or three leptons (d). The uncertainty bands include the statistical uncertainties for the background prediction as well as the 
systematic uncertainties for fake- or non-prompt-lepton backgrounds (using the matrix method) and charge-flip electrons. Not included
are theoretical uncertainties in the irreducible background contributions.
The rare category is defined in the text.}
\label{fig:Bkg_distribs} 
\end{figure}

Since the FNP lepton predictions from the MC template and matrix methods in the signal and validation regions are consistent 
with each other, a weighted average of the two results is used. 
With this approach, the combined estimate is always dominated 
by systematic uncertainties, which is not always the case when only the matrix method is used due to small number of events in the control regions.
To check the validity and robustness of the FNP lepton estimate, 
the distributions of several discriminating variables in data are compared 
with the predicted background after various requirements on the number of jets and $b$-jets. 
Examples of such distributions are shown in Figure~\ref{fig:Bkg_distribs}, 
and illustrate that the data are described by the prediction within uncertainties. The apparent disagreement 
for \meff\ above 1 TeV in Figure~\ref{fig:VRmeff2} is covered by the large theory uncertainty for the diboson background, which is not shown 
but amounts to about 30\% for \meff\ above 1 TeV.

\subsection{Validation of irreducible background estimates}
\label{sec:valid}

Dedicated validation regions are defined to verify the estimate of the $\ttbar V$, $WZ$ and $W^\pm W^\pm$ background 
in the signal regions. The corresponding selections are summarized in Table~\ref{tab:VRdef}. 
The overlap with the signal regions is resolved by removing events that are selected in the signal regions.  
The purity of the targeted background processes in these regions ranges from 35\% to 65\%. The expected signal contamination 
is generally below 5\% for models near the limit of exclusion in $\ttbar Z$, $WZ$ and $W^\pm W^\pm$ VRs and about 20\% in the $\ttbar W$ VR.
The observed yields, compared with the background predictions and uncertainties, 
are shown in Table~\ref{tab:VR_yields}. There is good agreement between data and the estimated background in all
the validation regions. 

\begin{table}[t!]
\hspace{0.5cm}
\def\arraystretch{1.1}
\centering
\resizebox{\textwidth}{!}
{\small
\begin{tabular}{|l|c|c|c|c|c|c|c|}
\hline    
Validation        &  $N_{\textrm{leptons}}^{\textrm{signal}}$    & $N_{b\textrm{-jets}}$  &  $N_{\textrm{jets}}$  & $\pt^{\textrm{jet}}$  & \met\ & \meff\  & Other \\
Region            &  &  &  & [GeV]  & [GeV] & [GeV]  & \\
\hline\hline
$\ttbar W$   	&$=2SS$     &$\geq 1$   & $\geq 4$ ($e^\pm e^\pm$, $e^\pm \mu^\pm$) & $>40$ & $> 45$  & $> 550$   & $\pt^{\ell_2}>40$~GeV\\
              	&           &       &  $\geq 3$ ($\mu^\pm \mu^\pm$)   &  $>25$ &      &          & $\sum \pt^{b\textrm{-jet}}/\sum \pt^{\textrm{jet}}>0.25$ \\ 
\hline
$\ttbar Z$    	&$\geq 3$  & $\geq 1$ & $\geq 3$ &  $>35$ &  --    & $> 450$  & $81<m_\text{SFOS}<101$~GeV \\
                &$\geq 1$ SFOS pair&     &          &       &         &         &  \\
\hline
$WZ$4j            & $=3$      &  $=0$ & $\geq 4$ &  $>25$   & --    & $> 450$ & $\met/\sum \pt^{\ell} < 0.7$ \\
\hline
$WZ$5j            & $=3$      &  $=0$ & $\geq 5$ &  $>25$   & --    & $> 450$ & $\met/\sum \pt^{\ell} < 0.7$  \\ 
\hline
$W^{\pm} W^{\pm}jj$ & $=2SS$      &  $=0$ & $\geq 2$ &   $>50$ & $> 55$  & $> 650$ & veto $81<\mee<101$~GeV  \\
               	  &               &	  &	     &         &   	&	  & $\pt^{\ell_2}>30$~GeV \\
               	  &		  &	  &	     &         &   	&	  & $\Delta R_\eta (\ell_{1,2},j)>0.7$  \\
               	  &		  &	  &	     &         &   	&	  & $\Delta R_\eta (\ell_1, \ell_2)>1.3$ \\
\hline
All VRs           & \multicolumn{7}{c|}{Veto events belonging to any SR} \\
\hline
\end{tabular}
}
\caption{Summary of the event selection in the validation regions (VRs). 
Requirements are placed on the number of signal leptons ($N_{\textrm{leptons}}^{\textrm{signal}}$), 
the number of $b$-jets with $\pt>\SI{20}{GeV}$ ($N_{b\textrm{-jets}}$) or the number of jets ($N_{\textrm{jets}}$) 
above a certain \pt threshold ($\pt^{\textrm{jet}}$). The two leading-\pt 
leptons are referred to as $\ell_{1,2}$ with decreasing \pt. Additional requirements are set 
on \met, \meff, the invariant mass of the two leading electrons \mee, the presence of SS 
leptons or a pair of same-flavour opposite-sign leptons (SFOS) and its invariant mass $m_\text{SFOS}$. 
A minimum angular separation between the leptons and the jets ($\Delta R_\eta (\ell_{1,2}, j)$) and between the two 
leptons ($\Delta R_\eta (\ell_{1}, \ell_2)$) is imposed in the $W^\pm W^\pm jj$ VR. 
For the two $WZ$ VRs the selection also relies on the ratio of the \met in the event to the sum of \pt of all signal leptons \pt (\met/$\sum{\pt^{\ell}}$). 
The ratio of the scalar sum of the \pt of all $b$-jets to that of all jets in the event 
($\sum \pt^{b{\textrm{-jet}}} / \sum{\pt^{\textrm{jet}}}$) is used in the $\ttbar W$ VR selection.}
\label{tab:VRdef}
\end{table}

\begin{table}[t!]
\hspace{0.5cm}
\def\arraystretch{1.1}
\centering
\begin{tabular}{|l|c|c|c|c|c|}
\hline    
 Validation Region       & $\ttbar W$           & $\ttbar Z$           & $WZ$4j              & $WZ$5j                 & $W^\pm W^{\pm}jj$     \\
\hline\hline
$\ttbar Z/\gamma^*$      & $ 6.2 \pm 0.9 $      & $ 123 \pm 17\hpO $   & $ 17.8 \pm 3.5\hpO$ & $ 10.1\pm 2.3\hpO $    & $ 1.06 \pm 0.22 $     \\
$\ttbar W$               & $ 19.0 \pm 2.9\hpO $ & $ 1.71 \pm 0.27 $    & $ 1.30 \pm 0.32$    & $ 0.45 \pm 0.14 $      & $ 4.1 \pm 0.8 $     \\
$\ttbar H$               & $  5.8 \pm 1.2 $     & $ 3.6 \pm 1.8 $      & $ 1.8 \pm 0.6$      & $ 0.96 \pm 0.34 $      & $ 0.69 \pm 0.14 $     \\
4$t$                     & $ 1.02 \pm 0.22 $    & $ 0.27 \pm 0.14 $    & $ 0.04 \pm 0.02$    & $ 0.03 \pm 0.02 $      & $ 0.03 \pm 0.02 $     \\
$W^{\pm}W^{\pm}$         & $ 0.5 \pm 0.4 $      &   --                 &   --                &   --                   & $ 26 \pm 14$     \\
$WZ$                     & $ 1.4 \pm 0.8 $      & $ 29 \pm 17 $        & $ 200 \pm 110$      & $ 70 \pm 40 $          & $ 27 \pm 14 $    \\
$ZZ$                     & $ 0.04 \pm 0.03 $    & $  5.5 \pm  3.1$     & $ 22 \pm 12$        & $  9\pm 5 $            & $ 0.53\pm 0.30 $     \\
Rare                     & $ 2.2 \pm 0.5 $      & $ 26 \pm 13 $        & $ 7.3 \pm 2.1$      & $  3.0 \pm 1.0 $       & $ 1.8 \pm 0.5 $     \\
Fake/non-prompt leptons  & $ 18 \pm 16$         & $ 22 \pm 14$         & $ 49 \pm 31 $       & $  17 \pm 12 $         & $ 13 \pm 10 $    \\
Charge-flip electrons    & $  3.4\pm 0.5 $      & --                   & --                  & --                     & $ 1.74 \pm 0.22 $     \\
\hline
Total SM background      & $ 57 \pm 16 $        & $212\pm 35\hpO$      & $300 \pm 130$       & $ 110 \pm 50\hpO $     & $ 77 \pm 31$     \\
\hline
Observed                 & $ 71 $               & $209$                & $257$               & $ 106 $                & $ 99  $            \\
\hline
\end{tabular}
\caption{The numbers of observed data and expected background events in the validation regions. 
The rare category is defined in the text. Background categories with yields shown as ``--'' 
do not contribute to a given region (e.g. charge flips in three-lepton regions) or their estimates are below 0.01 events. 
The displayed yields include all statistical and systematic uncertainties described in Section~\ref{sec:syst}.}
\label{tab:VR_yields}
\end{table}

\subsection{Systematic uncertainties}
\label{sec:syst}

Statistical uncertainties due to the number of data events in the loose and tight lepton control regions are considered in the FNP 
lepton background estimate. In the matrix method, the systematic uncertainties mainly come from potentially 
different compositions of $b$-jets, light-quark jets and photon conversions between the signal regions 
and the regions where the FNP lepton rates are measured. 
The uncertainty coming from the prompt-lepton contamination in the FNP lepton control regions is also considered.
Overall, the uncertainty in the FNP lepton rate $f$ amounts to 30\% at low $\pt$, and can reach 85\% for muons with $\pt>40$~GeV,
and 50\% for electrons with $\pt>20$~GeV; 
these values are driven respectively by the dependency of the isolation of non-prompt muons 
on the kinematic properties of the jets which emit them, 
and the uncertainty in the proportion of non-prompt electrons from heavy-flavoured hadron decays 
with respect to other sources of FNP electrons (mainly converted photons). 
The uncertainties in the prompt-lepton efficiency $\varepsilon$ are much smaller. 
The uncertainties in the FNP lepton background estimated with the matrix method in each VR and SR 
are then evaluated by propagating the $f$ and $\varepsilon$ uncertainties. 
In the MC template method, the systematic uncertainty is obtained by
changing the generator from \POWHEGBOX to \SHERPA and propagating uncertainties from the control region fit to the global
normalization scale factors applied to the MC samples. The uncertainties in these scale factors are in the range 75--80\%,
depending on the SRs.
When combining the results of the MC template method and the matrix method to obtain the final estimate, systematic uncertainties are propagated assuming 
conservatively a full correlation between the two methods. 

The uncertainty in the electron charge-flip probability mainly originates from the number of events in the regions used
in the charge-flip probability measurement and the uncertainty related to the background subtraction 
from the $Z$ boson's mass peak. The relative error in the charge-flip rate is below 20\% (30\%) for signal (candidate) 
electrons with \pt above 20 GeV.

The systematic uncertainties related to the estimated background from same-sign prompt leptons 
arise from the experimental uncertainties (jet energy scale calibration, 
jet energy resolution and $b$-tagging efficiency) as well as theoretical modelling and theoretical cross-section uncertainties.
The statistical uncertainty of the simulated event samples is also taken into account.

The cross-sections used to normalize the MC samples are varied according to the uncertainty in the 
cross-section calculation, which is 13\% for $\ttbar W$, 12\% for $\ttbar Z$ production~\cite{YR4}, 6\% for diboson
production~\cite{ATL-PHYS-PUB-2016-002}, 8\% for $\ttbar H$~\cite{YR4} and 30\% for 4$t$~\cite{Alwall:2014hca}. 
Additional uncertainties are assigned to some of these backgrounds to account for the theoretical modelling of the kinematic 
distributions in the MC simulation. For $\ttbar W$ and $\ttbar Z$, the predictions from the \AMCATNLO and \SHERPA generators are compared, 
and the renormalization and factorization scales used to generate these samples are varied independently within a factor of two, 
leading to a 15--35\% uncertainty in the expected SR yields for these processes. 
For diboson production, uncertainties are estimated by varying the QCD and matching scales, as well as the parton shower recoil scheme, 
leading to a 30--40\% uncertainty for these processes after the SR selections. 
For $\ttbar H$, 4$t$ and rare production processes, a 50\% uncertainty 
in their total contribution is assigned.

\section{Results and interpretation}
\label{sec:result}

\begin{figure}[t]
\begin{center}
\begin{subfigure}[t]{0.98\textwidth}\includegraphics[width=\textwidth]{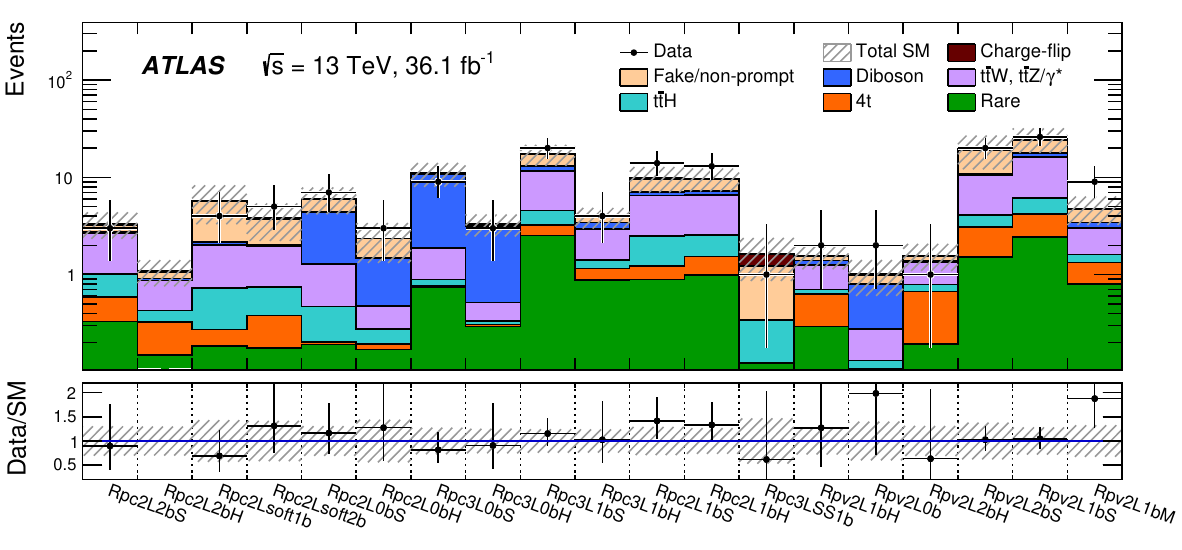}\caption{}\label{fig:Results_SRSum}\end{subfigure}
\begin{subfigure}[t]{1.08\textwidth}\includegraphics[width=\textwidth]{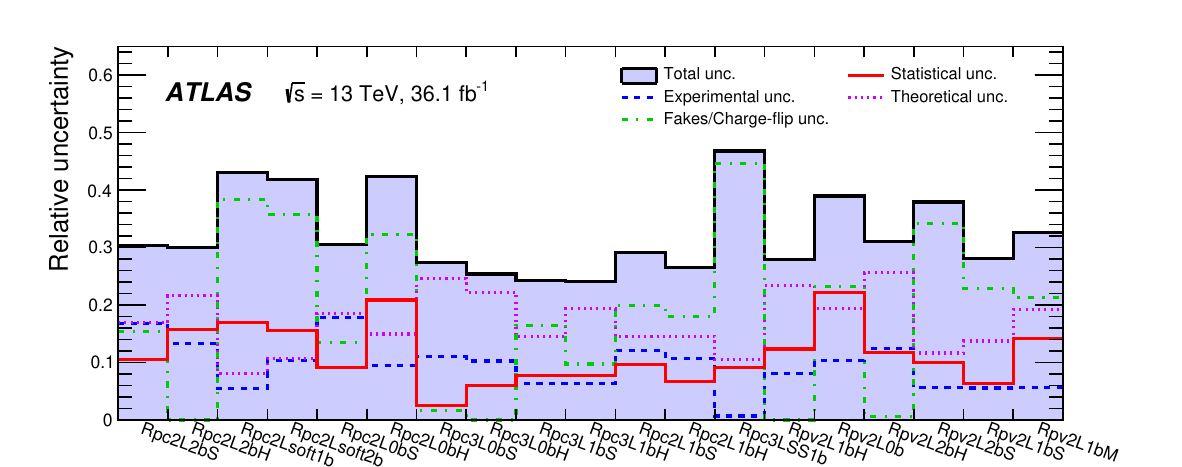}\caption{}\label{fig:Results_SystSum}\end{subfigure}
\end{center}
\caption{Comparison of (a) the observed and expected event yields in each signal region and (b) the relative uncertainties in the total 
background yield estimate. For the latter, ``statistical uncertainty'' corresponds to reducible and irreducible background 
statistical uncertainties. The background predictions correspond to those presented in Table~\ref{tab:SR_yields} and the 
rare category is explained in the text. } 
\label{fig:PlotSR}
\end{figure}

Figure~\ref{fig:Results_SRSum} shows the event yields for data and the expected background contributions 
in all signal regions. Detailed information about the yields can be found in Table~\ref{tab:SR_yields}.
In all 19 SRs the number of observed data events is consistent with the expected background within the uncertainties. 
The contributions listed in the rare category are dominated by triboson, $tWZ$ and $\ttbar WW$ 
production\footnote{Contributions from $WH$, $ZH$, $tZ$ and $\ttbar t$ production never represent more than 20\% of the rare background.} : 
the triboson processes generally dominate in the SRs with no $b$-jets, while $tWZ$ and $\ttbar WW$
dominate in the SRs with one and two $b$-jets, respectively.

\begin{table}
\scriptsize
\begin{center}
\vspace*{-0.035\textwidth}
\begin{tabular}{|l|c|c|c|c|c|c|}
\hline
Signal Region                  & \textbf{Rpc2L2bS} & \textbf{Rpc2L2bH}    & \textbf{Rpc2Lsoft1b} &\textbf{Rpc2Lsoft2b}& \textbf{Rpc2L0bS} & \textbf{Rpc2L0bH}\\
\hline
\hline
$\ttbar W$, $\ttbar Z\gamma^*$ & $1.6\pm0.4$       & $0.44\pm0.14$     & $1.3\pm0.4$       & $1.21\pm0.33$       & $0.82\pm0.31$       & $0.20\pm0.10$  \\
$\ttbar H$                     & $0.43\pm0.25$     & $0.10\pm0.06$     & $0.45\pm0.24$     & $0.36\pm0.21$       & $0.27\pm0.15$       & $0.08\pm0.07$   \\
4$t$                           & $0.26\pm0.13$     & $0.18\pm0.09$     & $0.09\pm0.05$     & $0.21\pm0.11$       & $0.01\pm0.01$       & $0.02\pm0.02$   \\
Diboson                        & $0.10\pm0.10$     & $0.04\pm0.02$     & $0.17\pm0.09$     & $0.05\pm0.03$       & $3.1\pm1.4$         & $1.0\pm0.5$   \\
Rare                           & $0.33\pm0.18$     & $0.15\pm0.09$     & $0.18\pm0.10$     & $0.17\pm0.10$       & $0.19\pm0.11$       & $0.17\pm0.10$   \\
Fake/non-prompt leptons        & $0.5\pm0.6$       & $0.15\pm0.15$     & $3.5\pm2.4$       & $1.7\pm1.5$         & $1.6\pm1.0$         & $0.9\pm0.9$   \\
Charge-flip electrons          & $0.10\pm0.01$     & $0.02\pm0.01$     & $0.08\pm0.02$     & $0.08\pm0.02$       & $0.05\pm0.01$       & $0.01\pm0.01$   \\ 
\hline
Total Background               & $3.3\pm1.0$       & $1.08\pm0.32$     & $5.8\pm2.5$       & $3.8 \pm1.6$        & $6.0\pm1.8$         & $2.4\pm1.0$   \\
\hline
Observed                       & $3$               & $0$               & $4$               & $5$                 & $7$                 & $3$   \\
\hline\hline
$S_{\textrm{obs}}^{95}$        & \ral{$5.5$}               & \ral{$3.6$}               & \ral{$6.3$}               & \ral{$7.7$}               & \ral{$8.3$}               & \ral{$6.1$}  \\
$S_{\textrm{exp}}^{95}$        & \ral{$5.6_{-1.5}^{+2.2}$} & \ral{$3.9_{-0.4}^{+1.4}$} & \ral{$7.1_{-1.5}^{+2.5}$} & \ral{$6.2_{-1.5}^{+2.6}$} & \ral{$7.5_{-1.8}^{+2.6}$} & \ral{$5.3_{-1.3}^{+2.1}$} \\
$\sigma_{\textrm{vis}}$ [fb]   & \ral{$0.15$}              & \ral{$0.10$}              & \ral{$0.17$}              & \ral{$0.21$}              & \ral{$0.23$}              & \ral{$0.17$}  \\
$p_{0}$ ($\textrm{Z}$)         & \ral{$0.71$ (--)}         & \ral{$0.91$ (--)}         & \ral{$0.69$ (--)}         & \ral{$0.30\ (0.5\sigma)$} & \ral{$0.36\ (0.4\sigma)$} & \ral{$0.35\ (0.4\sigma)$}  \\
\hline 
\end{tabular}

\vspace*{1cm}

\begin{tabular}{|l|c|c|c|c|c|c|c|}
\hline
Signal Region 		& \textbf{Rpc3L0bS } 	& \textbf{Rpc3L0bH } 	& \textbf{Rpc3L1bS } 	& \textbf{Rpc3L1bH } 	& \textbf{Rpc2L1bS } 	& \textbf{Rpc2L1bH } 	& \textbf{Rpc3LSS1b }\\
\hline
\hline
$\ttbar W$, $\ttbar Z\gamma^*$   & $0.98\pm0.25$       	& $0.18\pm0.08$       	& $7.1\pm1.1$       	& $1.54\pm0.28$       	& $4.0\pm1.0$ 		& $4.0\pm0.9$	  	&  --     		\\
$\ttbar H$              & $0.12\pm0.08$       	& $0.03\pm0.02$       	& $1.4\pm0.7$       	& $0.25\pm0.14$       	& $1.3\pm0.7$ 		& $1.0\pm0.6$	  	& $0.22\pm0.12$    	\\
4$t$	   		& $0.02\pm0.01$	  & $0.01\pm0.01$	  & $0.7\pm0.4$ 	  & $0.28\pm0.15$	  & $0.34\pm0.17$	  & $0.54\pm0.28$	  &  -- 		  \\
Diboson                  & $8.9\pm2.9$       	& $2.6\pm0.8$       	& $1.4\pm0.5$       	& $0.48\pm0.17$       	& $0.5\pm0.3$ 		& $0.7\pm0.3$ 		&  --     		\\
Rare                     & $0.7\pm0.4$       	& $0.29\pm0.16$       	& $2.5\pm1.3$       	& $0.9\pm0.5$       	& $0.9\pm0.5$		& $1.0\pm0.6$		& $0.12\pm0.07$    	\\
Fake/non-prompt leptons  & $0.23\pm0.23$       	& $0.15\pm0.15$       	& $4.2\pm3.1$       	& $0.5\pm0.5$       	& $2.5\pm2.2$ 		& $2.3\pm1.9$  		& $0.9\pm0.7$    	\\
Charge-flip electrons    &  --   		&  --    		&  --    		&  --     		& $0.25\pm0.04$ 	& $0.25\pm0.05$  	& $0.39\pm0.08$		\\	
\hline
Total Background         & $11.0\pm3.0\hpO$	       & $3.3\pm0.8$       	& $17\pm4\hpO$       	& $3.9\pm0.9$       	& $9.8\pm2.9$ 		& $9.8\pm2.6$  		& $1.6\pm0.8$	   	\\
\hline
Observed                 & $9$       		& $3$       		& $20$		       	& $4$       		& $14$  		&  $13$    		&  $1$			  \\
\hline\hline
$S_{\textrm{obs}}^{95}$       & \ral{$8.3$}	  	& $5.4$	   		& \ral{$14.7$}	    	& \ral{$6.1$}	     		& \ral{$13.7$}  		& \ral{$12.4$}   		& \ral{$3.9$}     		\\
$S_{\textrm{exp}}^{95}$       & \ral{$9.3_{-2.3}^{+3.1}$}	& \ral{$5.5_{-1.5}^{+2.2}$}	& \ral{$12.6_{-3.4}^{+5.1}$} 	& \ral{$5.9_{-1.8}^{+2.2}$}	& \ral{$10.0_{-2.6}^{+3.7}$}	& \ral{$9.7_{-2.6}^{+3.4}$}   & \ral{$4.0_{-0.3}^{+1.8}$}    \\
$\sigma_{\textrm{vis}}$ [fb] & \ral{$0.23$}		& \ral{$0.15$}  		& \ral{$0.41$}      		& \ral{$0.17$}      		& \ral{$0.38$}  		& \ral{$0.34$}   		& \ral{$0.11$}     		\\
$p_{0}$ ($\textrm{Z}$)        & \ral{$0.72$ (--)}  	& \ral{$0.85$ (--)}  		& \ral{$0.32\ (0.5\sigma)$}  & \ral{$0.46\ (0.1\sigma)$}  	& \ral{$0.17\ (1.0\sigma)$}  	& \ral{$0.21\ (0.8\sigma)$}	& \ral{$0.56$ (--)}	\\
\hline 
\end{tabular}

\vspace*{1cm}

\begin{tabular}{|l|c|c|c|c|c|c|}
\hline
Signal Region 		& \textbf{Rpv2L1bH } 	& \textbf{Rpv2L0b } 	& \textbf{Rpv2L2bH } 	& \textbf{Rpv2L2bS } 	& \textbf{Rpv2L1bS }  	& \textbf{Rpv2L1bM  } \\
\hline
\hline
$\ttbar W$, $\ttbar Z\gamma^*$   & $0.56\pm0.14$       	& $0.14\pm0.08$       	& $0.56\pm0.15$       	& $6.5\pm1.3$       	& $10.1\pm1.7\hpO$  	& $1.4\pm0.5$       \\
$\ttbar H$              & $0.07\pm0.05$       	& $0.02\pm0.02$       	& $0.12\pm0.07$       	& $1.0\pm0.5$       	& $1.9\pm1.0$  		& $0.28\pm0.15$      \\
4$t$       		 & $0.34\pm0.17$       	& $0.01\pm0.01$       	& $0.48\pm0.24$       	& $1.6\pm0.8$       	& $1.8\pm0.9$  		& $0.53\pm0.27$      \\
Diboson                  & $0.14\pm0.06$       	& $0.52\pm0.21$       	& $0.04\pm0.02$       	& $0.42\pm0.16$       	& $1.7\pm0.6$  		& $0.42\pm0.15$      \\
Rare                     & $0.29\pm0.17$       	& $0.10\pm0.06$       	& $0.19\pm0.13$   	& $1.5\pm0.8$       	& $2.4\pm1.2$  		& $0.8\pm0.4$      \\
Fake/non-prompt leptons  & $0.15\pm0.15$       	& $0.18\pm0.31$		& $0.15\pm0.15$       	& $8\pm7$       	& $6\pm6$  		& $1.3\pm1.2$      \\
Charge-flip electrons    & $0.02\pm0.01$       	& $0.03\pm0.02$       	& $0.03\pm0.01$       	& $0.46\pm0.08$       	& $0.74\pm0.12$  	& $0.10\pm0.02$      \\        
\hline
Total Background         & $1.6\pm0.4$       	& $1.0\pm0.4$	       	& $1.6\pm0.5$       	& $19\pm7\hpO$       	& $25\pm7\hpO$  		& $4.8\pm1.6$       \\
\hline
Observed                 & $2$       		& $2$		        & $1$       		& $20$		        & $26$  		& $9$       		\\
\hline\hline
$S_{\textrm{obs}}^{95}$       & \ral{$4.8$}	  	& \ral{$5.2$}   		& \ral{$3.9$}	    		& \ral{$17.5$}	     	& \ral{$18.1$}  		& \ral{$11.4$}	 	\\
$S_{\textrm{exp}}^{95}$       & \ral{$4.1_{-0.4}^{+1.9}$}	& \ral{$4.0_{-0.3}^{+1.7}$}	& \ral{$4.1_{-0.4}^{+1.8}$}	& \ral{$16.8_{-4.2}^{+5.2}$}	& \ral{$17.2_{-4.2}^{+5.9}$}  & \ral{$7.3_{-1.8}^{+2.5}$}	 \\
$\sigma_{\textrm{vis}}$ [fb] & \ral{$0.13$}       	& \ral{$0.14$}       		& \ral{$0.11$}       		& \ral{$0.48$}       		& \ral{$0.50$}  		& \ral{$0.31$}      		\\
$p_{0}$ ($Z$)        & \ral{$0.33\ (0.4\sigma)$}   & \ral{$0.19\ (0.9\sigma)$}	& \ral{$0.55$ (--)}    	& \ral{$0.48\ (0.1\sigma)$}  & \ral{$0.44\ (0.2\sigma)$}  & \ral{$0.07\ (1.5\sigma)$}     	\\
\hline 
\end{tabular}

\vspace*{1cm}

\vspace*{-0.01\textheight}\caption{Numbers of events observed in the signal regions compared with the expected backgrounds. 
The rare category is defined in the text. Background categories with yields shown as a ``--'' 
do not contribute to a given region (e.g. charge flips in three-lepton regions) or their estimates are below 0.01. 
The 95\% confidence level (CL) upper limits are shown on the observed and expected numbers of BSM events, $S_{\textrm{obs}}^{95}$ and $S_{\textrm{exp}}^{95}$ 
(as well as the $\pm 1\sigma$ excursions from the expected limit), respectively. The 95\% CL upper limits on the visible cross-section 
($\sigma_{\textrm{vis}}$) are also given. Finally, the p-values (p$_0$) give the probabilities to observe a deviation 
from the predicted background at least as large as that in the data. The number of equivalent Gaussian standard deviations ($Z$) is also 
shown when $p_{0}<0.5$.}
\label{tab:SR_yields}
\end{center}
\end{table}

Figure~\ref{fig:Results_SystSum} summarizes the contributions from the different sources of systematic uncertainty 
to the total SM background predictions in the signal regions. The uncertainties amount to 25--50\% of the 
total background depending on the signal region, dominated by systematic uncertainties coming from the reducible background or the theory. 

In the absence of any significant deviation from the SM predictions, upper limits on possible BSM contributions to the signal regions are derived, 
as well as exclusion limits on the masses of SUSY particles in the benchmark scenarios of Figure~\ref{fig:feynman}. 
The HistFitter framework~\cite{Baak:2014wma}, which utilizes a profile-likelihood-ratio test~\cite{Cowan:2010js}, 
is used to establish 95\% confidence intervals using the CL$_\mathrm{s}$ prescription~\cite{Read:2002hq}. 
The likelihood is built as the product of a Poisson probability density function describing the observed number of events in the signal region 
and, to constrain the nuisance parameters associated with the systematic uncertainties, 
Gaussian distributions whose widths correspond to the sizes of these uncertainties; 
Poisson distributions are used instead for MC simulation statistical uncertainties.
Correlations of a given nuisance parameter between the backgrounds and the signal are taken into account when relevant. 
The hypothesis tests are performed for each of the signal regions independently. 

Table~\ref{tab:SR_yields} presents 95\% confidence level (CL) observed (expected) model-independent upper limits 
on the number of BSM events, $S_{\textrm{obs}}^{95}$ ($S_{\textrm{exp}}^{95}$), that may contribute to the signal regions. 
Normalizing these by the integrated luminosity $L$ of the data sample, they can be interpreted as upper limits on the visible 
BSM cross-section ($\sigma_{\textrm{vis}}$), defined as $\sigma_{\textrm{vis}}=\sigma_{\textrm{prod}}\times A \times\epsilon=S_{\textrm{obs}}^{95}/L$, where 
$\sigma_{\textrm{prod}}$ is the production cross-section, $A$ the acceptance and $\epsilon$ the reconstruction efficiency. The largest 
deviation of the data from the background prediction corresponds to an excess of 1.5 standard deviations in the Rpv2L1bM SR.

\begin{figure}[p]
\centering
\begin{subfigure}[t]{0.38\textwidth}\includegraphics[width=\textwidth]{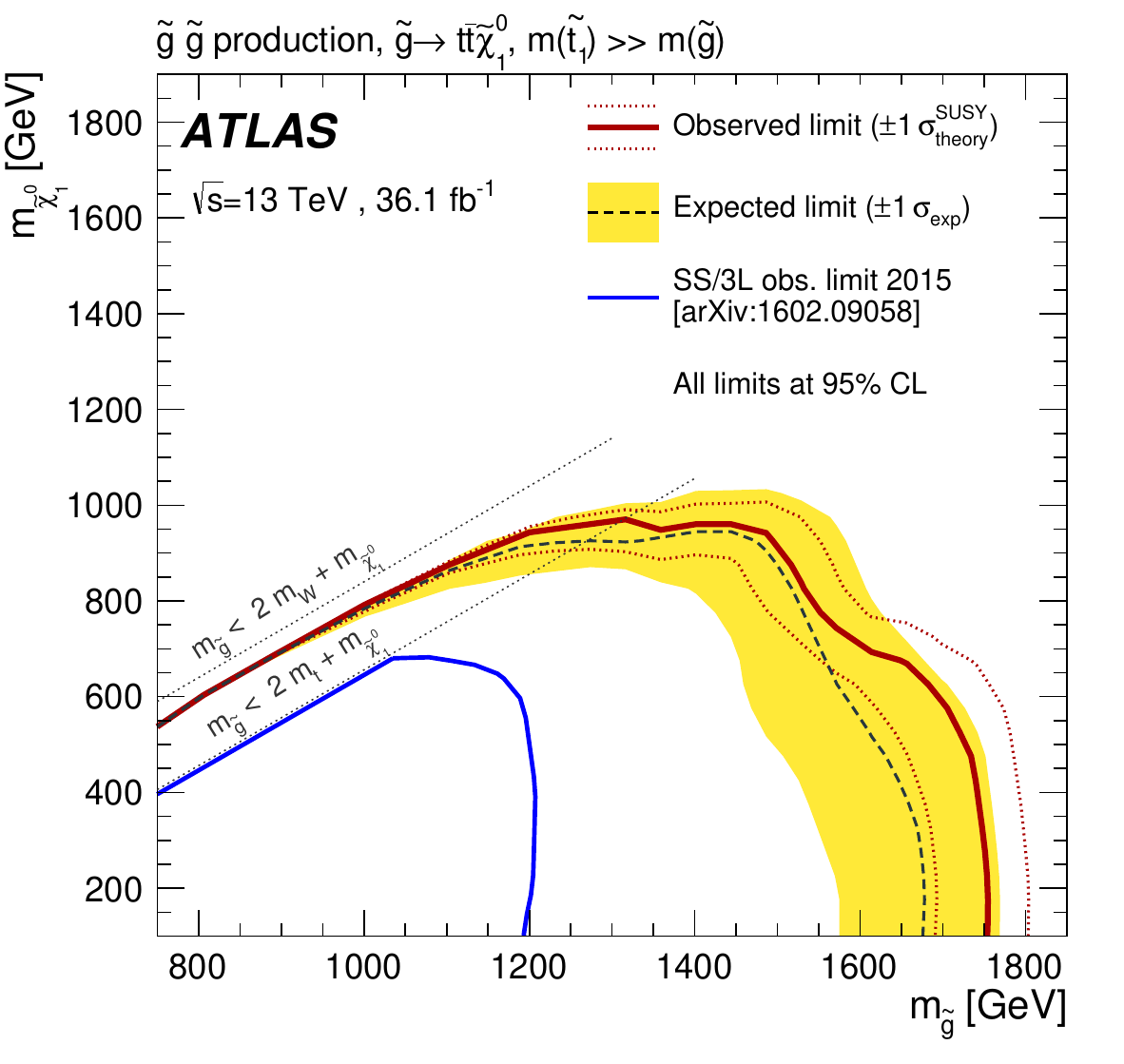}\caption{Rpc2L2bS/H, Rpc2Lsoft1b/2b}\label{fig:limits_feynman_gtt}\end{subfigure}
\begin{subfigure}[t]{0.38\textwidth}\includegraphics[width=\textwidth]{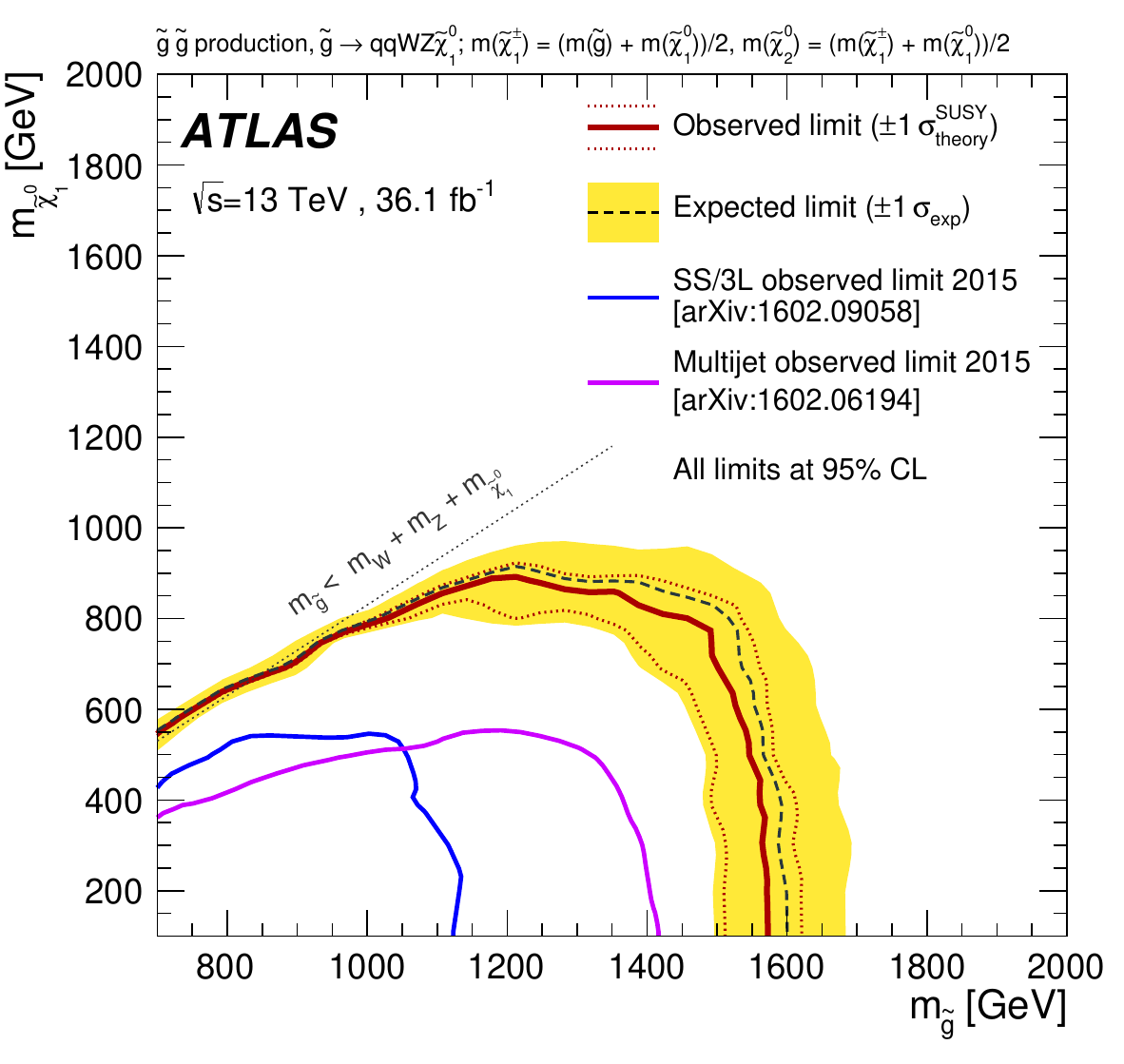}\caption{Rpc2L0bS, Rpc2L0bH}\label{fig:limits_feynman_gg2WZ}\end{subfigure}
\begin{subfigure}[t]{0.38\textwidth}\includegraphics[width=\textwidth]{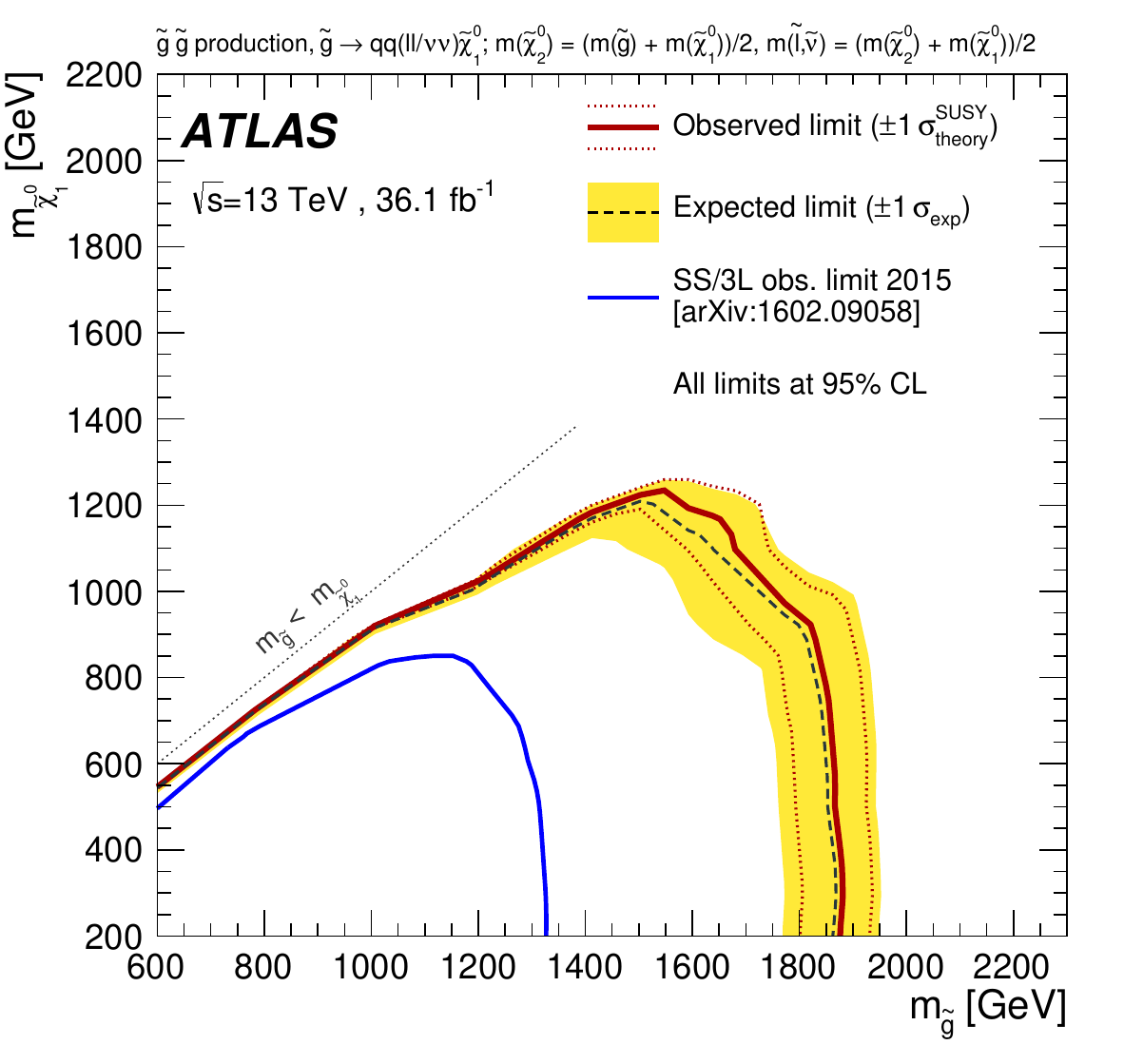}\caption{Rpc3L0bS, Rpc3L0bH}\label{fig:limits_feynman_gg2sl}\end{subfigure}
\begin{subfigure}[t]{0.38\textwidth}\includegraphics[width=\textwidth]{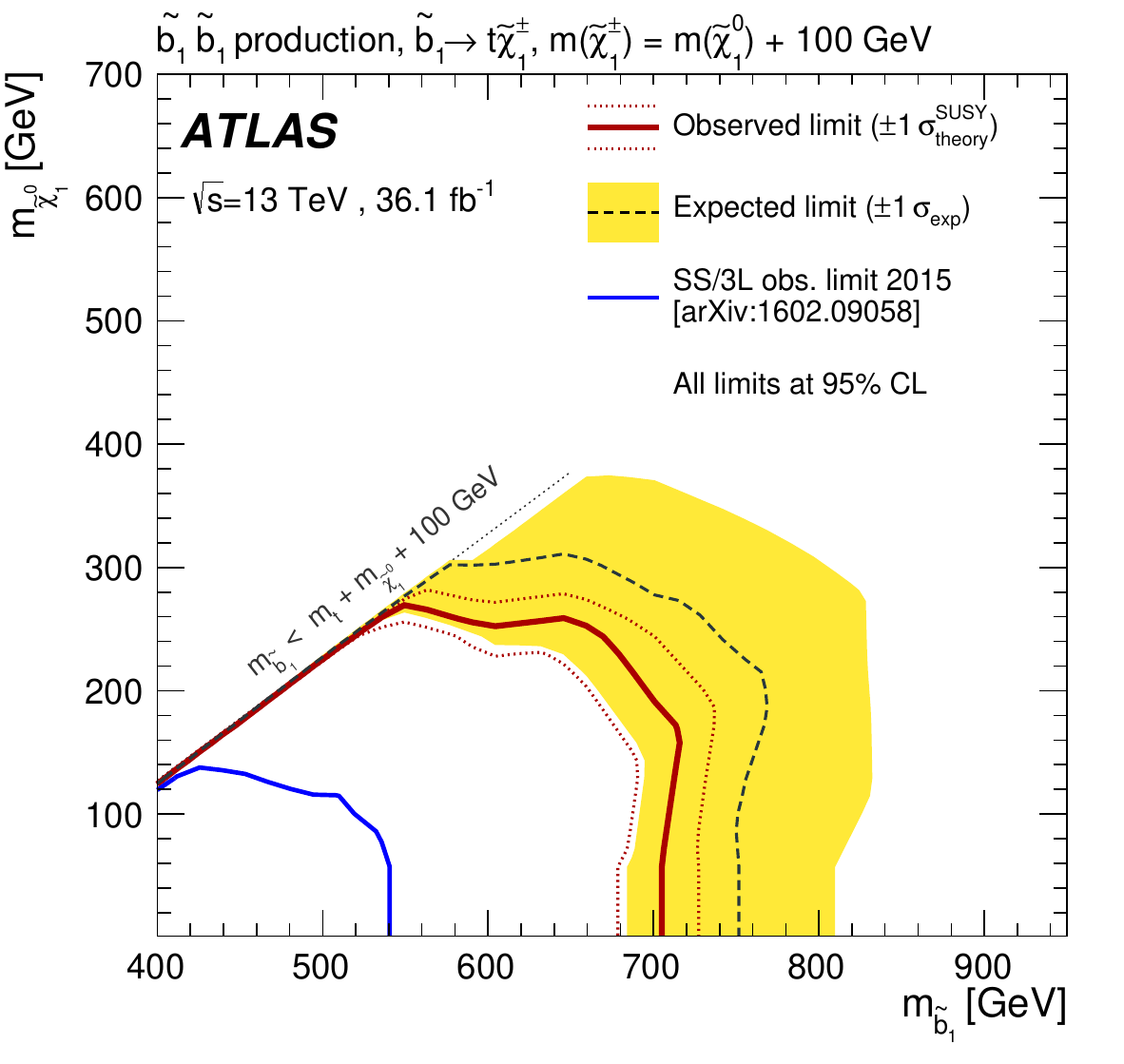}\caption{Rpc2L1bS, Rpc2L1bH}\label{fig:limits_feynman_b1b1}\end{subfigure}
\begin{subfigure}[t]{0.38\textwidth}\includegraphics[width=\textwidth]{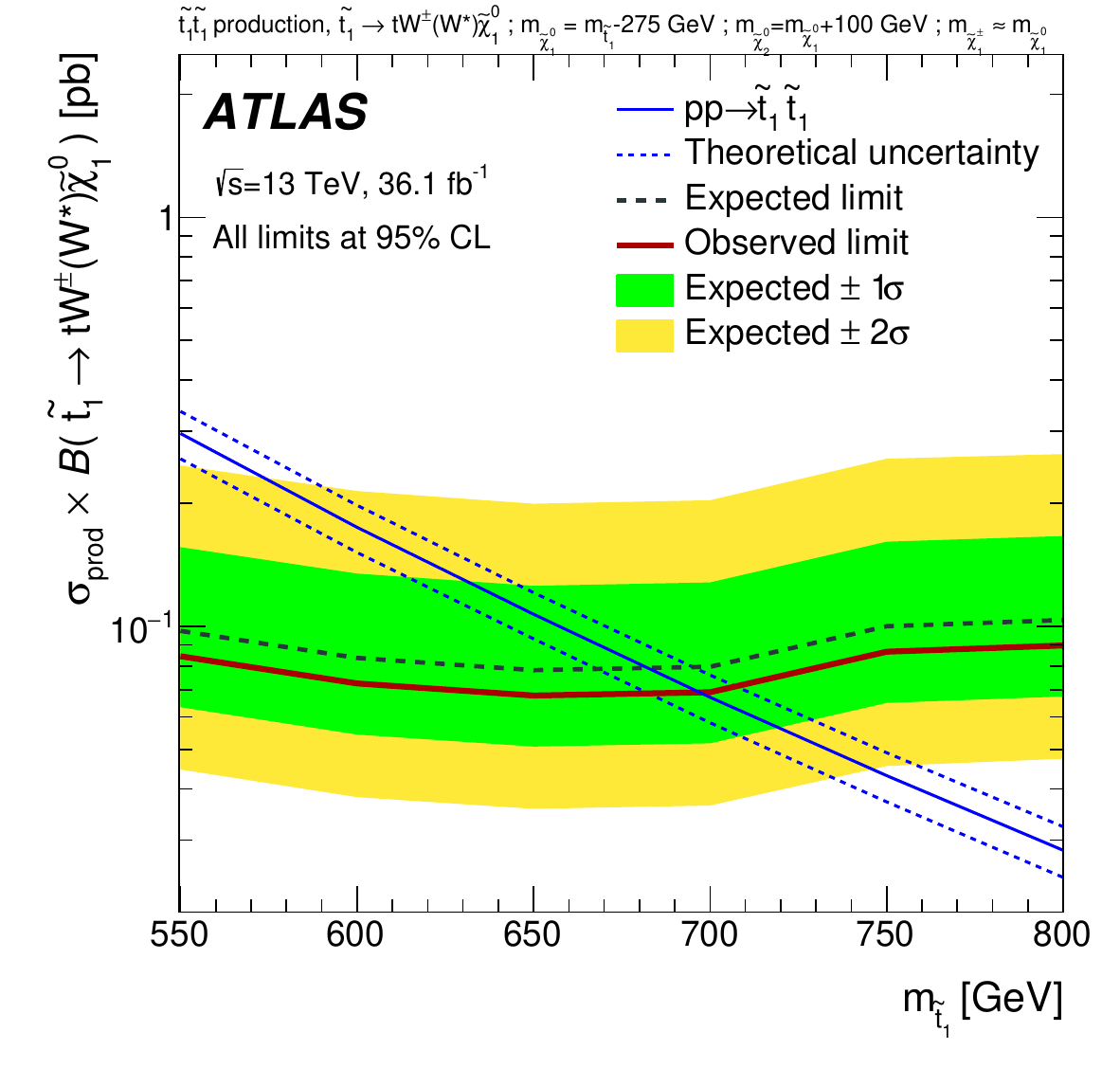}\caption{Rpc3LSS1b}\label{fig:limits_feynman_t1t1}\end{subfigure}
\caption{Observed and expected exclusion limits on the $\tilde{g}$, \sbottomone, \stopone\ and \ninoone masses 
in the context of RPC SUSY scenarios with simplified mass spectra. The signal regions used to obtain the limits are specified in the subtitle of 
each scenario. All limits are computed at 95\% CL. The dotted lines around the observed
limit illustrate the change in the observed limit as the nominal signal cross-section is scaled up and down
by the theoretical uncertainty. The contours of the band around the expected 
limit are the $\pm$1$\sigma$ results ($\pm$2$\sigma$ is also considered in Figure~(e), 
including all uncertainties except the theoretical uncertainties in the signal cross-section. In Figures~(a)--(d), 
the diagonal line indicates the kinematic limit for the decays in each specified scenario and results are compared with the observed limits obtained 
by previous ATLAS searches~\cite{paperSS3L,Aad:2016jxj}.}
\label{fig:Results_Limits_RPC} 
\end{figure} 

\begin{figure}[p]
\centering
\begin{subfigure}[t]{0.38\textwidth}\includegraphics[width=\textwidth]{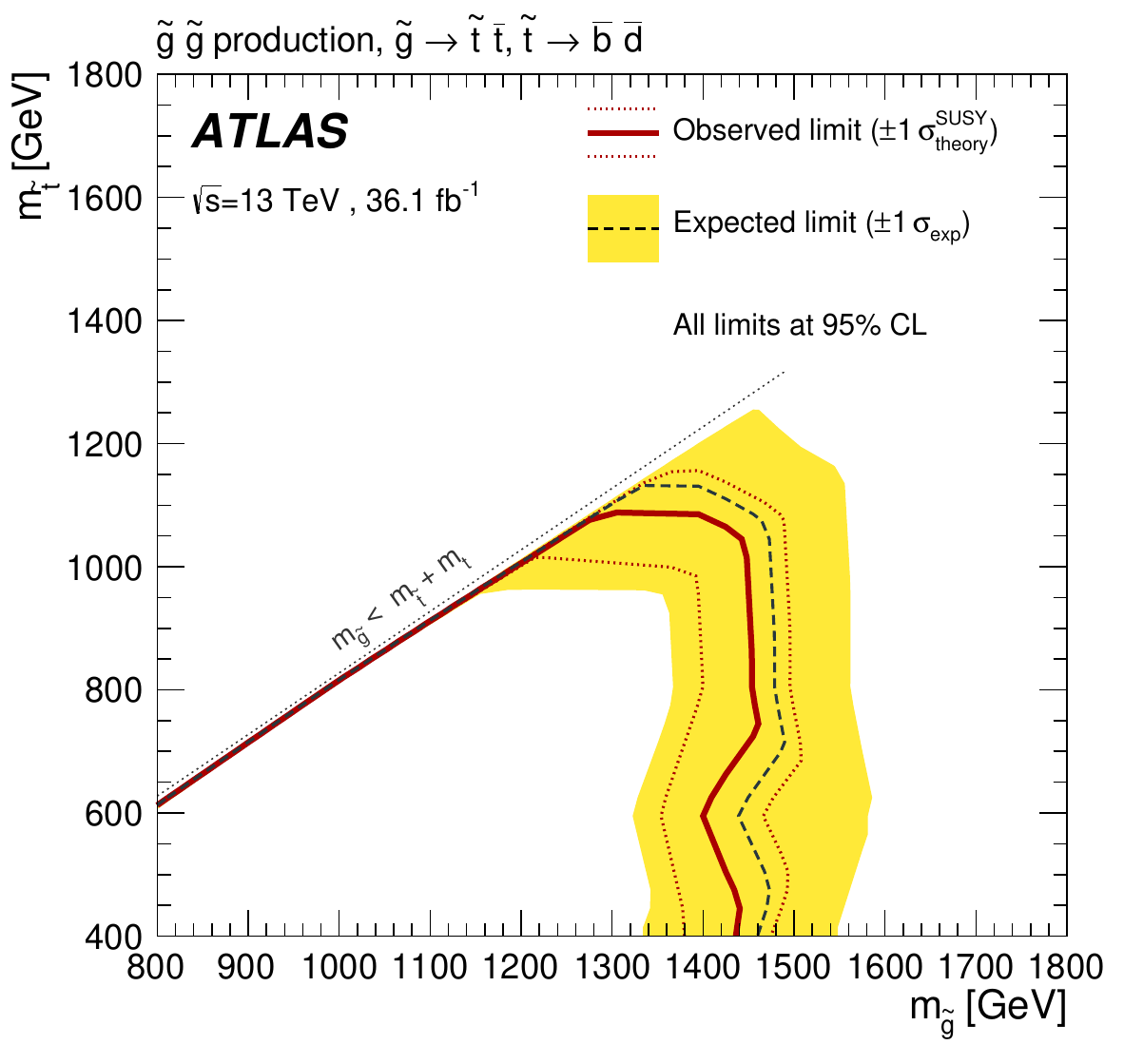}\caption{Rpv2L1bH}\label{fig:limits_feynm_rpv_gl313}\end{subfigure}
\begin{subfigure}[t]{0.38\textwidth}\includegraphics[width=\textwidth]{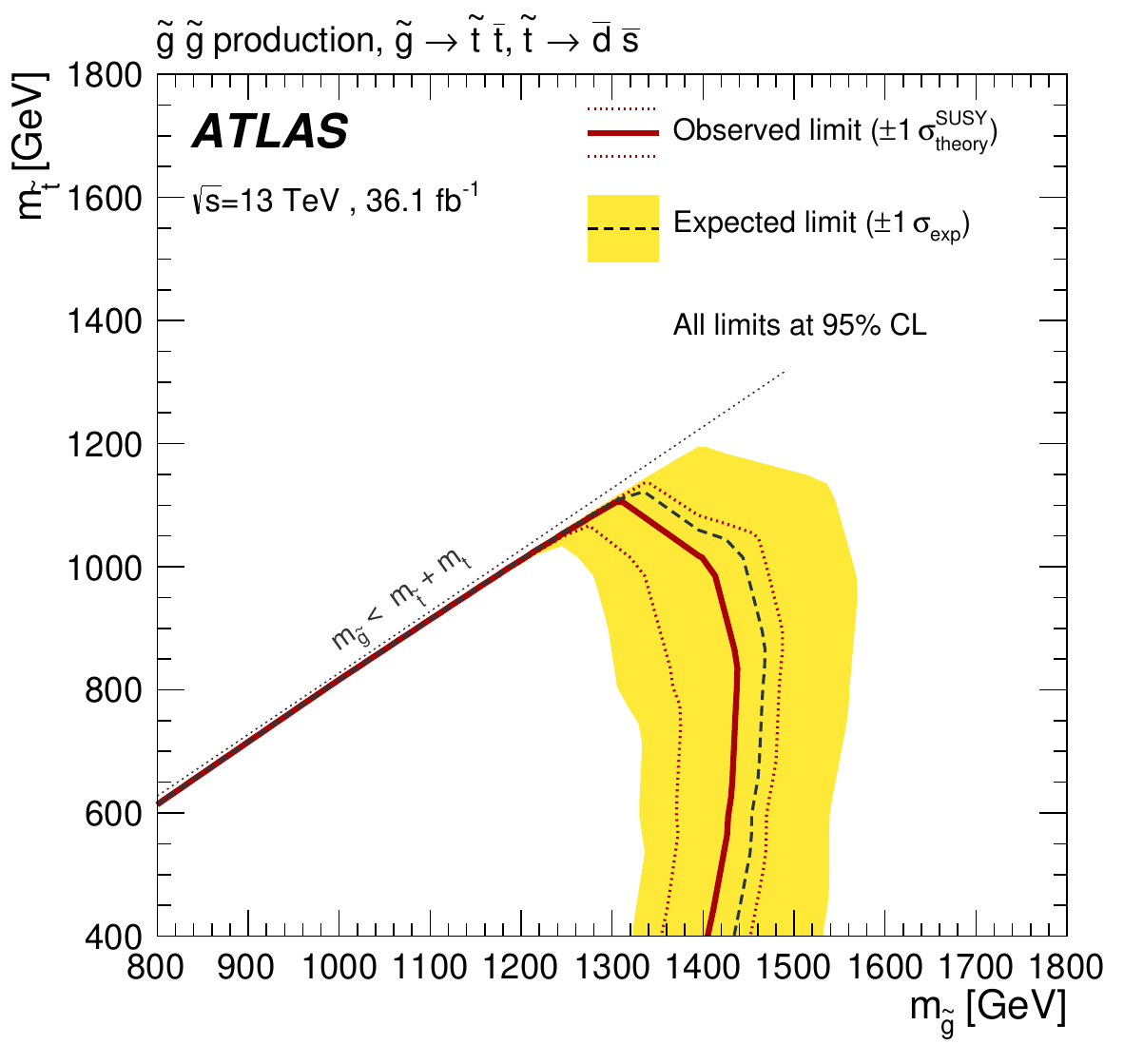}\caption{Rpv2L1bH}\label{fig:limits_feynm_rpv_gl321}\end{subfigure}
\begin{subfigure}[t]{0.38\textwidth}\includegraphics[width=\textwidth]{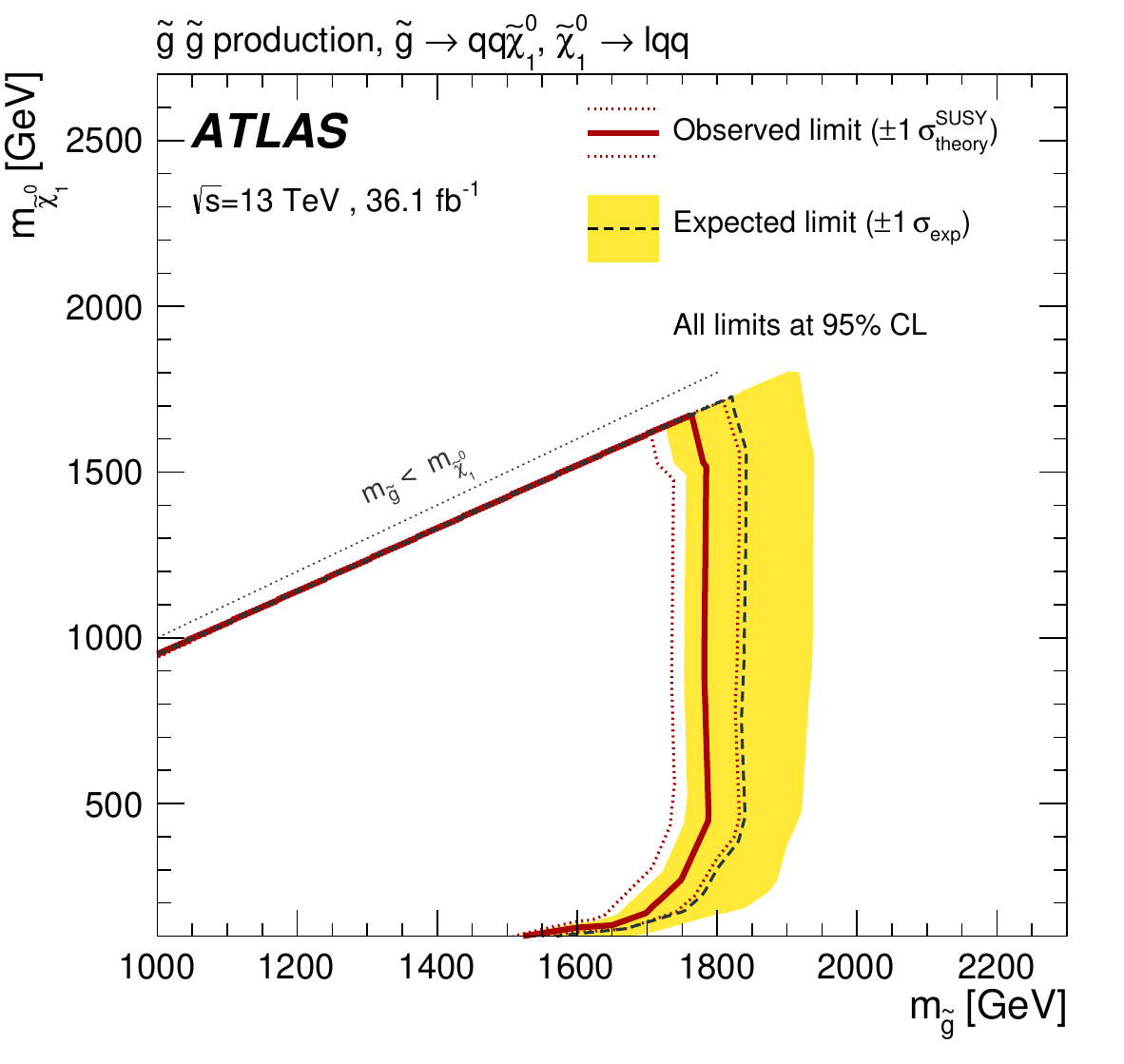}\caption{Rpv2L0b}\label{fig:limits_feynm_rpv_glprime}\end{subfigure}
\begin{subfigure}[t]{0.38\textwidth}\includegraphics[width=\textwidth]{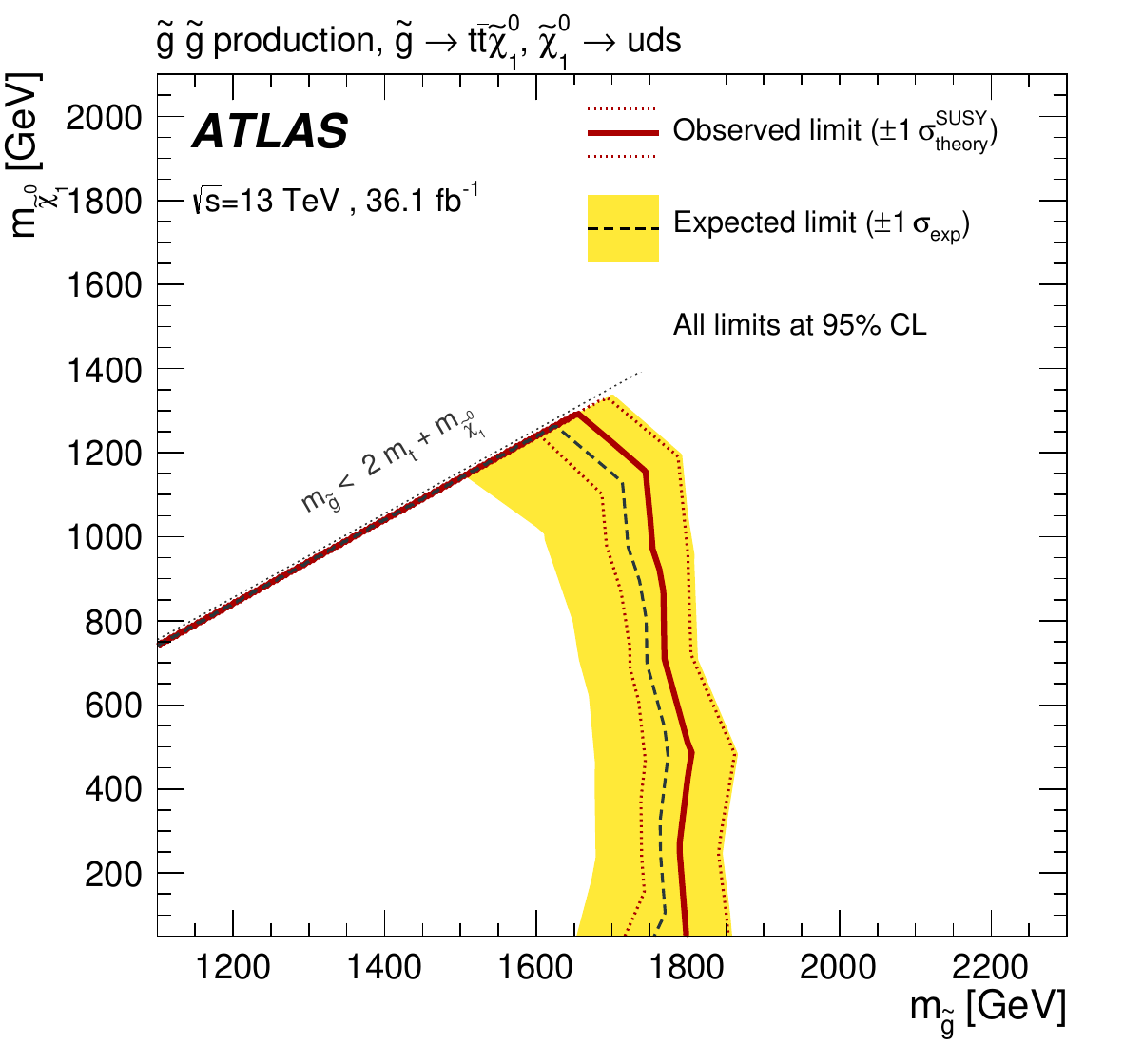}\caption{Rpv2L2bH}\label{fig:limits_feynm_rpv_gl112l}\end{subfigure}
\begin{subfigure}[t]{0.38\textwidth}\includegraphics[width=\textwidth]{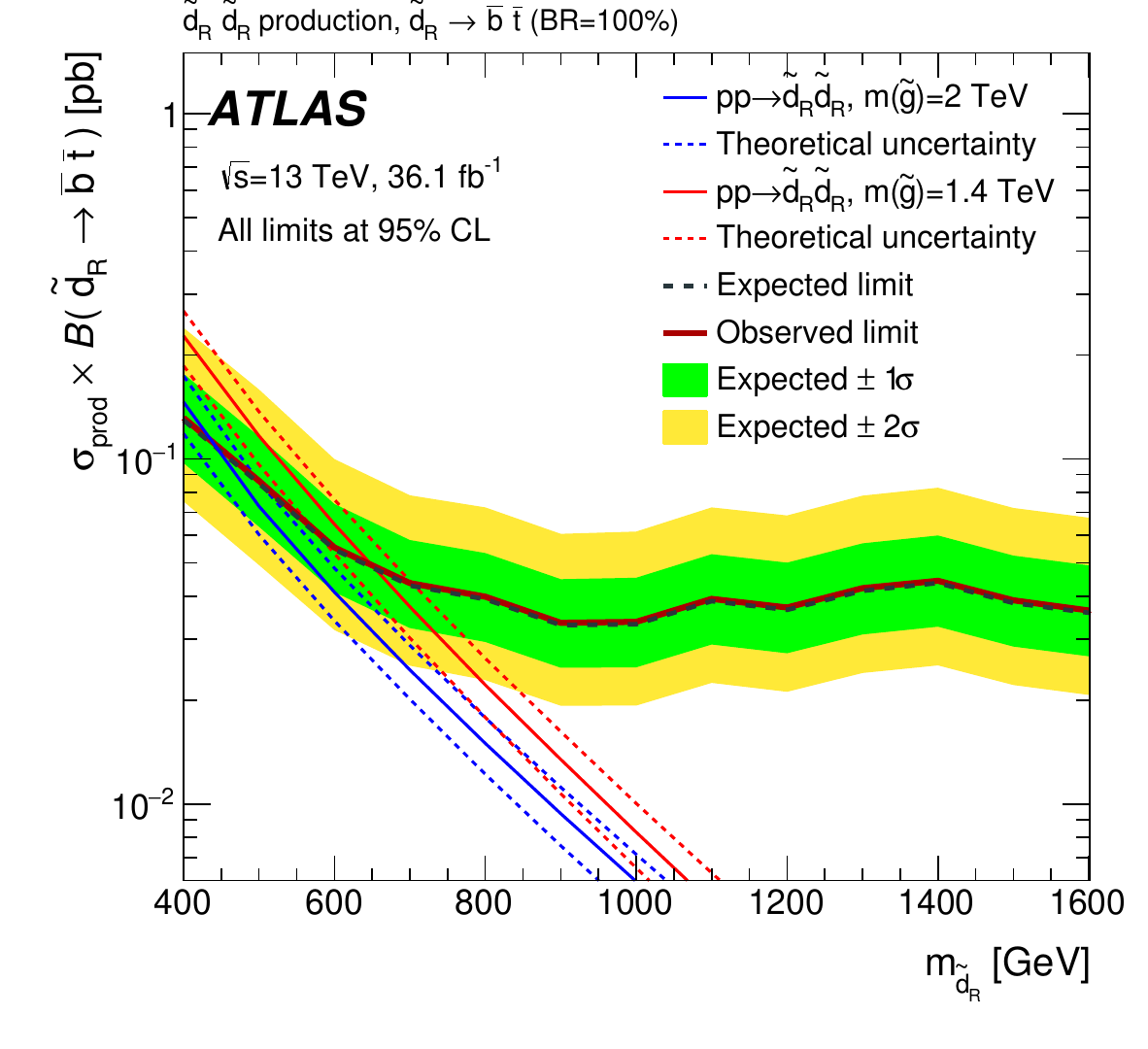}\caption{Rpv2L2bS}\label{fig:limits_feynm_rpv_sd313}\end{subfigure}
\begin{subfigure}[t]{0.38\textwidth}\includegraphics[width=\textwidth]{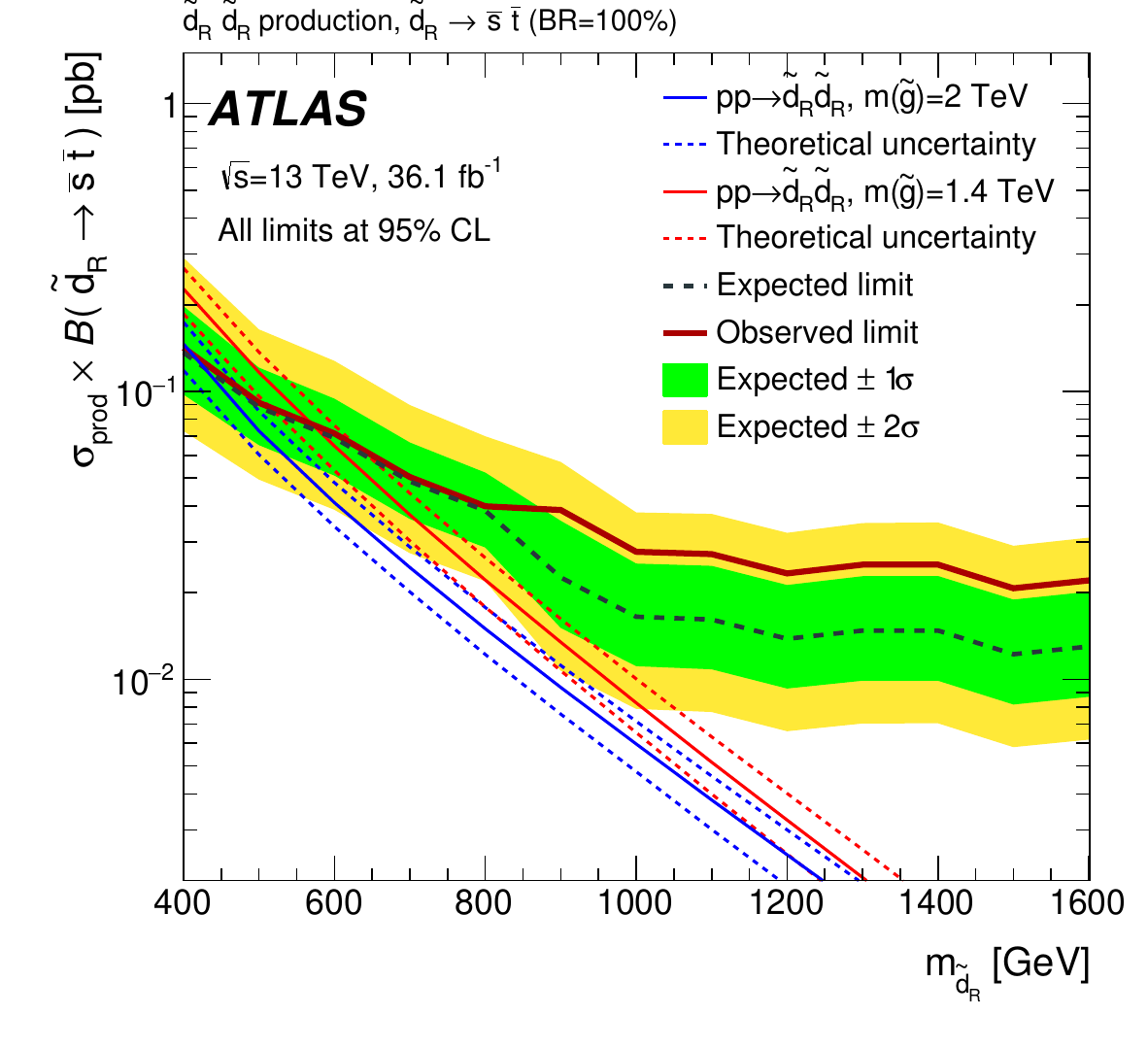}\caption{Rpv2L1bS, Rpv2L1bM}\label{fig:limits_feynm_rpv_sd321}\end{subfigure}
\caption{
Observed and expected exclusion limits on the $\tilde{g}$, \stopone, $\tilde d_\mathrm{R}$ and \ninoone masses 
in the context of RPV SUSY scenarios with simplified mass spectra 
featuring $\gluino\gluino$ or $\tilde d_\mathrm{R}\tilde d_\mathrm{R}$ pair production with exclusive decay modes. 
The signal regions used to obtain the limits are specified in the subtitle of each scenario. All limits are computed at 95\% CL. 
The dotted lines around the observed limit illustrate the change in the observed limit as the nominal signal cross-section is scaled up and down
by the theoretical uncertainty. The contours of the band around the expected limit are the $\pm$1$\sigma$ results, 
including all uncertainties except theoretical uncertainties in the signal cross-section ($\pm$2$\sigma$ is also considered in 
Figures~\ref{fig:limits_feynm_rpv_sd313} and \ref{fig:limits_feynm_rpv_sd321}). In Figures~\ref{fig:limits_feynm_rpv_gl313}--\ref{fig:limits_feynm_rpv_gl112l}, 
the diagonal line indicates the kinematic limit for the decays in each specified scenario. For Figures~\ref{fig:limits_feynm_rpv_sd313} 
and \ref{fig:limits_feynm_rpv_sd321}, theoretical production cross-sections are shown for two different gluino masses in red (1.4 TeV) and
blue (2.0 TeV).}
\label{fig:Results_Limits_RPV} 
\end{figure}

\begin{figure}[t]
\centering
\includegraphics[width=0.50\textwidth]{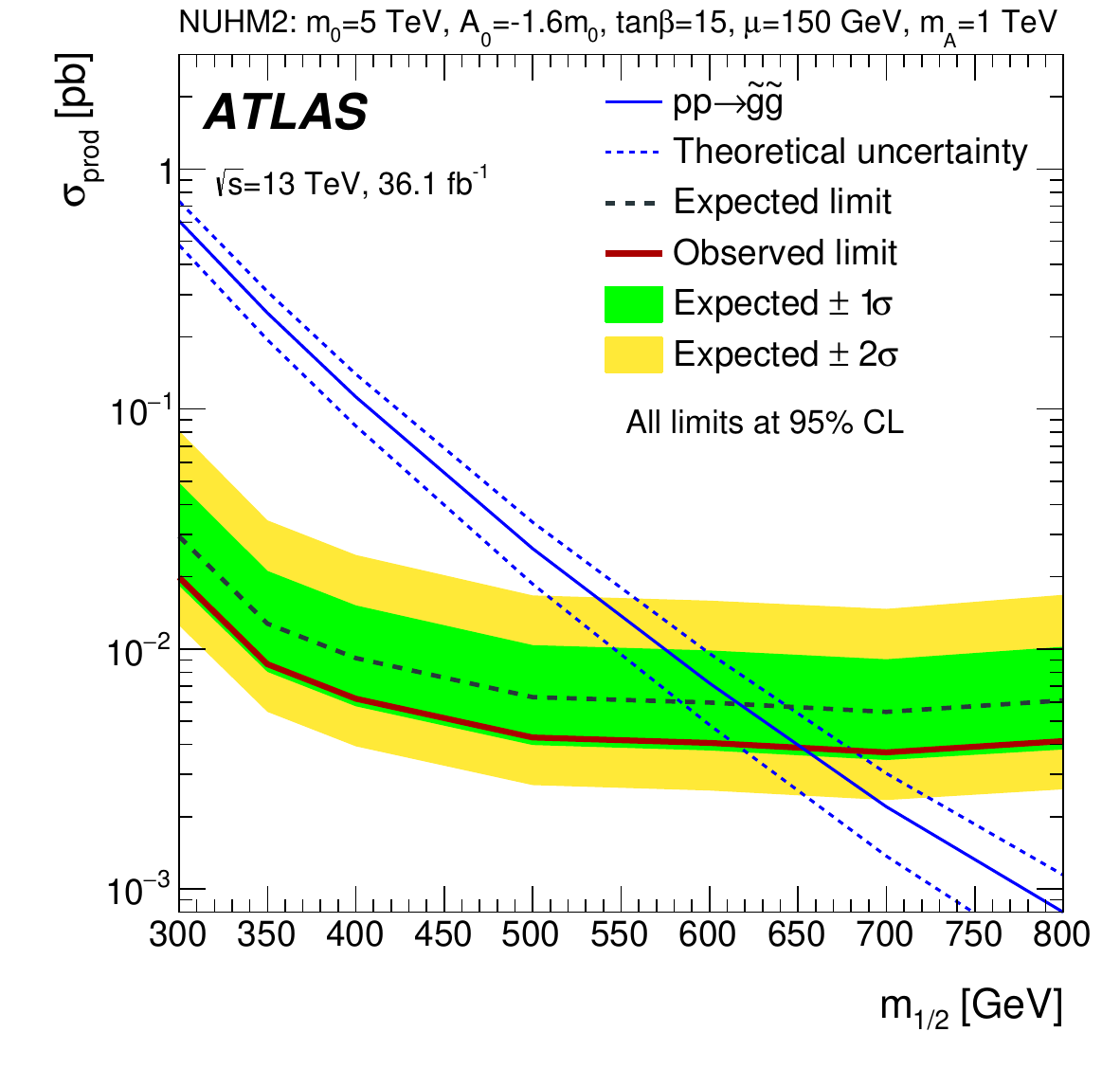}
\caption{Observed and expected exclusion limits as a function of $m_{1/2}$ in the NUHM2 model~\cite{Ellis:2002iu,Ellis:2002wv}.
The signal region Rpc2L2bH is used to obtain the limits. 
The contours of the green (yellow) band around the expected limit are the $\pm$1$\sigma$ ($\pm$2$\sigma$) results, including all uncertainties. The limits are computed at 95\% CL.}
\label{fig:Results_Limits_NUHM2} 
\end{figure} 

Exclusion limits at 95\% CL are also set on the masses of the superpartners involved in the SUSY benchmark scenarios considered. 
Apart from the NUHM2 model, simplified models are used, corresponding to a single production mode and with 100\% branching ratio to a specific decay chain, 
with the masses of the SUSY particles not involved in the process set to very high values. 
Figures~\ref{fig:Results_Limits_RPC}, \ref{fig:Results_Limits_RPV} and \ref{fig:Results_Limits_NUHM2} show the exclusion limits in all 
the models considered in Figure~\ref{fig:feynman} and the NUHM2 model. The assumptions about the decay chain considered for the different SUSY particles are 
stated above each figure. For each region of the signal parameter space, the SR with the best expected sensitivity is chosen.

For the RPC models, the limits set are compared with the existing limits set by other ATLAS SUSY 
searches~\cite{paperSS3L,Aad:2016jxj}. For the models shown in Figure~\ref{fig:Results_Limits_RPC}, 
the mass limits on gluinos and bottom squarks are up to 400~GeV higher than the previous limits, reflecting the improvements 
in the signal region definitions as well as the increase in integrated luminosity. Gluinos with masses up to 1.75~TeV
are excluded in scenarios with a light $\ninoone$ in Figure~\ref{fig:limits_feynman_gtt}. This limit is extended to 1.87~TeV when 
$\ninotwo$ and slepton masses are in-between the gluino and the $\ninoone$ masses (Figure~\ref{fig:limits_feynman_gg2sl}). More generally, gluino masses 
below 1.57~TeV and bottom squarks with masses below 700 GeV 
are excluded in models with a massless LSP. The ``compressed'' regions, where SUSY particle masses are close to each other, are also better covered 
and LSP masses up to 1200 and 250~GeV are excluded in the gluino and bottom squark pair-production models, respectively. Of particular
interest is the observed exclusion of models producing gluino pairs with an off-shell top quark in the decay (Figure~\ref{fig:feynman_gttOffshell}), 
see Figure~\ref{fig:limits_feynman_gtt}. In this case, models are excluded for mass differences between the gluino and neutralino of 205 GeV (only ~35 GeV
larger than the minimum mass difference for decays into two on-shell $W$ bosons and two $b$-quarks) for a gluino mass below 0.9
TeV. The Rpc3LSS1b SR allows the exclusion of top squarks with masses below 700~GeV when the top squark decays to a top quark and a cascade of electroweakinos 
$\ninotwo \to \chinoonepm W^{\mp} \to W^{*} W^{\mp} \ninoone$ (see Figure~\ref{fig:limits_feynman_t1t1} for the conditions 
on the sparticle masses).

For the RPV models with gluino pair production (Figures~\ref{fig:limits_feynm_rpv_gl313} -- \ref{fig:limits_feynm_rpv_gl112l}), 
a generic exclusion of gluinos with masses below 1.3 TeV is obtained. Weaker exclusion limits, typically around 500 GeV, 
are obtained in models with pair production of $\tilde d_\mathrm{R}$ (Figures~\ref{fig:limits_feynm_rpv_sd313},~\ref{fig:limits_feynm_rpv_sd321}). 

Finally, in the NUHM2 model with low fine-tuning, values of the parameter $m_{1/2}$ below 615 GeV are excluded, 
corresponding to gluino masses below 1500 GeV (Figure~\ref{fig:Results_Limits_NUHM2}).

\FloatBarrier

\section{Conclusion}
\label{sec:conclusion}

A search for supersymmetry in events with two same-sign leptons or at least three leptons, multiple jets, 
$b$-jets and large $\met$ and/or large $\meff$ is presented. 
The analysis is performed with proton--proton collision data at $\sqrt{s}=13$~TeV 
collected in 2015 and 2016 with the ATLAS detector at the Large Hadron Collider 
corresponding to an integrated luminosity of 36.1 fb$^{-1}$. 
With no significant excess over the Standard Model prediction observed,
results are interpreted in the framework of simplified models featuring gluino 
and squark production in $R$-parity-conserving and $R$-parity-violating scenarios. Lower limits on particle 
masses are derived at 95\% confidence level. 
In the $\gluino\gluino$ simplified RPC models considered, gluinos with masses up to 1.87~TeV
are excluded in scenarios with a light $\ninoone$. RPC models with bottom squark masses below 700~GeV
are also excluded in a $\sbottomone\sbottomonebar$ simplified model with $\sbottomone\to tW^-\ninoone$ and a light $\ninoone$. 
In RPV scenarios, masses of down squark-rights are probed up to $m_{\tilde d_\mathrm{R}}\approx$ 500~GeV. 
All models with gluino masses below 1.3 TeV are excluded, greatly extending the previous exclusion limits obtained within this search.
Model-independent limits on the cross-section of a possible signal contribution to the signal regions are set.

\section*{Acknowledgements}


We thank CERN for the very successful operation of the LHC, as well as the
support staff from our institutions without whom ATLAS could not be
operated efficiently.

We acknowledge the support of ANPCyT, Argentina; YerPhI, Armenia; ARC, Australia; BMWFW and FWF, Austria; ANAS, Azerbaijan; SSTC, Belarus; CNPq and FAPESP, Brazil; NSERC, NRC and CFI, Canada; CERN; CONICYT, Chile; CAS, MOST and NSFC, China; COLCIENCIAS, Colombia; MSMT CR, MPO CR and VSC CR, Czech Republic; DNRF and DNSRC, Denmark; IN2P3-CNRS, CEA-DSM/IRFU, France; SRNSF, Georgia; BMBF, HGF, and MPG, Germany; GSRT, Greece; RGC, Hong Kong SAR, China; ISF, I-CORE and Benoziyo Center, Israel; INFN, Italy; MEXT and JSPS, Japan; CNRST, Morocco; NWO, Netherlands; RCN, Norway; MNiSW and NCN, Poland; FCT, Portugal; MNE/IFA, Romania; MES of Russia and NRC KI, Russian Federation; JINR; MESTD, Serbia; MSSR, Slovakia; ARRS and MIZ\v{S}, Slovenia; DST/NRF, South Africa; MINECO, Spain; SRC and Wallenberg Foundation, Sweden; SERI, SNSF and Cantons of Bern and Geneva, Switzerland; MOST, Taiwan; TAEK, Turkey; STFC, United Kingdom; DOE and NSF, United States of America. In addition, individual groups and members have received support from BCKDF, the Canada Council, CANARIE, CRC, Compute Canada, FQRNT, and the Ontario Innovation Trust, Canada; EPLANET, ERC, ERDF, FP7, Horizon 2020 and Marie Sk{\l}odowska-Curie Actions, European Union; Investissements d'Avenir Labex and Idex, ANR, R{\'e}gion Auvergne and Fondation Partager le Savoir, France; DFG and AvH Foundation, Germany; Herakleitos, Thales and Aristeia programmes co-financed by EU-ESF and the Greek NSRF; BSF, GIF and Minerva, Israel; BRF, Norway; CERCA Programme Generalitat de Catalunya, Generalitat Valenciana, Spain; the Royal Society and Leverhulme Trust, United Kingdom.

The crucial computing support from all WLCG partners is acknowledged gratefully, in particular from CERN, the ATLAS Tier-1 facilities at TRIUMF (Canada), NDGF (Denmark, Norway, Sweden), CC-IN2P3 (France), KIT/GridKA (Germany), INFN-CNAF (Italy), NL-T1 (Netherlands), PIC (Spain), ASGC (Taiwan), RAL (UK) and BNL (USA), the Tier-2 facilities worldwide and large non-WLCG resource providers. Major contributors of computing resources are listed in Ref.~\cite{ATL-GEN-PUB-2016-002}.

\printbibliography

\newpage 
\begin{flushleft}
{\Large The ATLAS Collaboration}

\bigskip

M.~Aaboud$^\textrm{137d}$,
G.~Aad$^\textrm{88}$,
B.~Abbott$^\textrm{115}$,
O.~Abdinov$^\textrm{12}$$^{,*}$,
B.~Abeloos$^\textrm{119}$,
S.H.~Abidi$^\textrm{161}$,
O.S.~AbouZeid$^\textrm{139}$,
N.L.~Abraham$^\textrm{151}$,
H.~Abramowicz$^\textrm{155}$,
H.~Abreu$^\textrm{154}$
% ...
\bigskip
\\
% ...
$^{12}$ Institute of Physics, Azerbaijan Academy of Sciences, Baku, Azerbaijan\\
% ...
$^{88}$ CPPM, Aix-Marseille Universit{\'e} and CNRS/IN2P3, Marseille, France\\
% ...
$^{115}$ Homer L. Dodge Department of Physics and Astronomy, University of Oklahoma, Norman OK, United States of America\\
% ...
$^{119}$ LAL, Univ. Paris-Sud, CNRS/IN2P3, Universit{\'e} Paris-Saclay, Orsay, France\\
% ...
$^{137}$ $^{(a)}$ Facult{\'e} des Sciences Ain Chock, R{\'e}seau Universitaire de Physique des Hautes Energies - Universit{\'e} Hassan II, Casablanca; $^{(b)}$ Centre National de l'Energie des Sciences Techniques Nucleaires, Rabat; $^{(c)}$ Facult{\'e} des Sciences Semlalia, Universit{\'e} Cadi Ayyad, LPHEA-Marrakech; $^{(d)}$ Facult{\'e} des Sciences, Universit{\'e} Mohamed Premier and LPTPM, Oujda; $^{(e)}$ Facult{\'e} des sciences, Universit{\'e} Mohammed V, Rabat, Morocco\\
% ...
$^{139}$ Santa Cruz Institute for Particle Physics, University of California Santa Cruz, Santa Cruz CA, United States of America\\
% ...
$^{151}$ Department of Physics and Astronomy, University of Sussex, Brighton, United Kingdom\\
% ...
$^{154}$ Department of Physics, Technion: Israel Institute of Technology, Haifa, Israel\\
$^{155}$ Raymond and Beverly Sackler School of Physics and Astronomy, Tel Aviv University, Tel Aviv, Israel\\
% ...
$^{161}$ Department of Physics, University of Toronto, Toronto ON, Canada\\
% ...
$^{a}$ Also at Department of Physics, King's College London, London, United Kingdom\\
$^{b}$ Also at Institute of Physics, Azerbaijan Academy of Sciences, Baku, Azerbaijan\\
$^{c}$ Also at Novosibirsk State University, Novosibirsk, Russia\\
$^{d}$ Also at TRIUMF, Vancouver BC, Canada\\
$^{e}$ Also at Department of Physics {\&} Astronomy, University of Louisville, Louisville, KY, United States of America\\
% ...
$^{*}$ Deceased
\end{flushleft}




\end{document}